\documentclass[aps,prl,reprint,showpacs,showkeys]{revtex4-1}
\usepackage{microtype}
\usepackage{amsthm}
\usepackage{amsmath}
\usepackage{amsfonts}
\usepackage[mathscr]{eucal}
\usepackage{amssymb,epsfig,setspace}
\usepackage{graphicx}
\usepackage{caption}
\usepackage{subcaption}
\usepackage{upgreek}
\usepackage{graphicx}
\usepackage{epsfig}
\usepackage{psfrag}

\newcommand{\be}{\begin{equation}}
\newcommand{\ee}{\end{equation}}
\newcommand{\bc}{\begin{center}}
\newcommand{\ec}{\end{center}}
\newcommand{\ba}{\begin{array}}
\newcommand{\ea}{\end{array}}
\newcommand{\bea}{\begin{eqnarray}}
\newcommand{\eea}{\end{eqnarray}}

\newcommand{\lgl}{\mbox{\large{(}}}
\newcommand{\lgr}{\mbox{\large{)}}}

\newcommand{\sml}{\mbox{\small{(}}}
\newcommand{\smr}{\mbox{\small{)}}}

\newcommand{\tyl}{\mbox{\tiny{(}}}
\newcommand{\tyr}{\mbox{\tiny{)}}}

\begin{document}
% \preprint{CFT/Maestro-1}
\title[Thermalization in many-particle quantum walks]{Thermalization in many-particle quantum walks}%
\author{Dibwe Pierrot Musumbu}%
\email{pierrotmus@gmail.com} \affiliation{Center for Theoretical Physics of the Polish Academy of Science, Al. Lotnikow 32/46, 02-668 Warsaw, Poland }%
\author{Maria Przybylska}%
\email{M.Przybylska@if.uz.zgora.pl} \affiliation{Institute of Physics, University of Zielona G\'ora, Licealna 9, 65--417 Zielona G\'ora, Poland }%
\author{Andrzej J.~Maciejewski} \email{andrzej.j.maciejewski@gmail.com}
\affiliation{Institute of Astronomy, University of Zielona G\'ora, Licealna 9, PL-65--417 Zielona G\'ora, Poland.}%

\date{\today}%

\begin{abstract}
  Many-particles quantum walks of particles obeying Bose statistics moving on graphs of various topologies are introduced. A single coin tossing commands the conditional shift operation over the whole graph. Vertices particle densities, the mean values of the phase space variables, second order spatial correlation and counting statistics are evaluated and simulated. Evidence of an universal dynamics is presented. 
\end{abstract}
                      
\maketitle
\section{Introduction}
\label{sec:intro}
Classical random walks have proven to be a powerful  tool in both physics and mathematics through their numerous applications in 
algorithmics: Markov processes, Monte Carlo simulations, etc...Recent developments of quantum computation and quantum information raises 
renewed interests  directed toward advantages in quantum algorithms \cite{Aharonov:01,Nayak:01,Milburn:01,Childs:01} and aiming to adapt random walks to the quantum world.
 Problems such as bosons sampling \cite{Seshadreesan:01}, development 			
of quantum metrology, simulations of many-body quantum systems \cite{Knill:01}, waveguides arrays \cite{Peruzzo:01,Hamilton:01,DeNicola:01} and many others, have driven the newly developed quantum walks toward the use of many quantum walkers. 

In recent developments, the studies on quantum walks are also directed toward understanding the dynamics of many-particle systems on lattices 
\cite{laflamme:01,laflamme:02, Mayer01,Xue02,Sheridan:01}. In that regards, studying many-particles on graph implies exploring individual vertices dynamics and how much they affect the entire system. Furthermore, looking at the evolution of both individual vertices and the entire system enables to explore  how long the system can be confined in a certain regime. This is similar to measuring how fast the quantum walk spreads or how confined the quantum walk stays in a small neighborhood \cite{Aharonov:01}. 

In this work we study many-particle quantum walks on closed graphs: the cyclic graph, the double hexagon graph and the Petersen graph. We simulate discrete time quantum walks on these graphs for twelve bosons on ten vertices. We focus on the universality of dynamics appearing in the counting statistics and we analyze how vertices occupation numbers evolve in time. We observe the universal behavior  of the counting statistics that is independent of the types of graph and the initial conditions of the systems as long as the number of vertices of the graphs is the same and the number of quantum walkers remains unchanged. 
We recall that the quantum walk is a unitary process and therefore, its probability distribution and many other observables do not converge \cite{julia:02}. Nevertheless
 we must precise that a step in discrete time quantum walks is a two stage operation involving the state space and an additional degree of freedom known as the coin
degrees \cite{Childs04}. This puts a restriction on the unitarity in discrete quantum walks but it can be overcomes as suggested in \cite{Watrous:01}. Consequently, we do not expect the dynamics into our systems to converge. We control how long the quantum walks is confined on a vertex and how this affects the evolution of the entire quantum walks. We use the vertices particle distribution, the multimode phase space dynamics, the vertices counting statistics as well as the second order position correlation functions to study time evolution of our systems. In the counting statistics we
observed a change of regime. 
 
The title of this article introduce the concept of thermalization in many-particle quantum walks. 
In classical systems thermalization is associated with the time evolution of physical observables such as momentum, energy etc...
toward a Boltzmann distribution independently of theirs initial conditions. This behavior is explained by classical physics 
in the following way. Almost all particle trajectories independently of theirs initial conditions, quickly begin to look alike
because theirs dynamics is governed by nonlinear equations that drive them to explore the constant-energy manifold ergodically.
However, if the classical system possesses additional first integrals that are functionally independent of the hamiltonian and 
each other, then the phase curves of the system of particles are confined to a highly restricted region of
the energy manifold. Hence, the statistical predictions fail and the system does not thermalize \cite{Gallavotti:99}. But the time evolution of isolated quantum systems is linear  and theirs spectra are discrete \cite{Krylov:01}. Moreover in isolated quantum systems, the conditions directing conserved quantities to provide independent constraints on the relaxation dynamics are not well understood \cite{Rigol01,Rigol02,Rigol03}. Roughly speaking,
thermalization in isolated quantum systems can be understood as a time evolution to a state that belongs to a proper subspace of the Hilbert space of the system. The thermalization of a quantum dynamics corresponds with an observable that mean values behave according to the prediction of an appropriate statistical mechanical ensemble (represented by a certain density matrix). In our simulations
of the quantum walks on the three systems, we have observed such a behavior for the particles counting statistics. We need to precise that we are analyzing the quantum thermalization with respect to {\it the eigenstates thermalization hypothesis} \cite{Deutsch, Srednicki}. This means that the dimension of the fraction of the Hilbert space containing the thermalized state remains constant in time. We want to avoid the generalities by reserving a more detailed analysis in the conclusion.

This article is planned as follows. We start by introducing the mathematical formulation of quantum walks using identical particles in the spatial field representation in order to construct the graph many-particle state (GMP state). Next we construct the conditional shift operator needed for the implementation of the shared coin's many-particle quantum walks using the topology of the graph. After, we present the results of the simulations for many-particle quantum walks. In each case, we present the evolution of the vertices particles distributions, multimode phase space dynamics, the second order position correlation functions and the vertices counting statistics. The last section contains our conclusions and gives a short outlook.
 \vspace*{-0.75cm}
%==============================
\section{The graph many-particle state}
%==============================
%\vspace*{-0.3cm}
\begin{figure}[h!]
  \centering
      \includegraphics[width=0.47\textwidth]{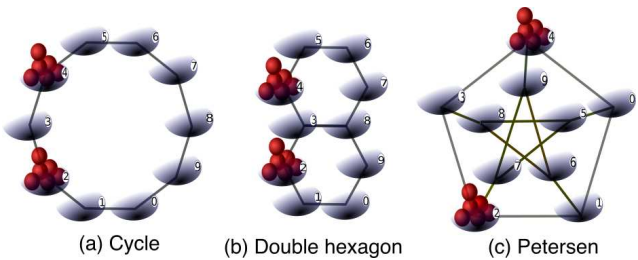}
  \caption{Systems of 12 bosons on 10 vertices graphs. \label{fig:123}}
%\vspace*{-0.6cm}
\end{figure}
Let us consider a family of operators $\widehat{\mathbf{\Uppsi}}^{\dagger}{\sml x\smr}$ acting on the vacuum  state $|\Omega\rangle$ such that they create position eigenvector $|x\rangle$:
\begin{equation}
 \label{2.1-1}
 \widehat{\mathbf{\Uppsi}}^{\dagger}{\sml x\smr}|\Omega\rangle=|x\rangle.
\end{equation}
This field operator creates a particle at position $x$. Similarly we have the family of operators $\widehat{\mathbf{\Uppsi}}{\sml x\smr}$ called the field annihilation operators defined as 
\begin{equation}
 \label{2.1-2}
 \widehat{\mathbf{\Uppsi}}{\sml x\smr}|x\rangle=|\Omega\rangle.
\end{equation}
In this work, we use discrete field operators therefore when considering the integration over space we will use summation over all the position $x$. We recall that the field operators satisfying a set of axioms were introduced by Jordan and Wigner \cite{Wigner01} in order to describe identical particles. Those axioms lead to two types of fields describing the two types of particles: bosons and fermions. The creation $\widehat{\mathbf{\Uppsi}}^{\dagger}{\sml x_{\mbox{\tiny{$2$}}}\smr}$ and the annihilation $\widehat{\mathbf{\Uppsi}}{\sml x_{\mbox{\tiny{$1$}}}\smr}$
 and the  field operator satisfy a commutator algebra, 
\begin{eqnarray}
 \label{2.1-1.1}
 [\widehat{\mathbf{\Uppsi}}{\sml x_{\mbox{\tiny{$1$}}}\smr},\,\widehat{\mathbf{\Uppsi}}^{\dagger}{\sml x_{\mbox{\tiny{$2$}}}\smr}]&=&\delta(x_{\mbox{\tiny{$1$}}}-x_{\mbox{\tiny{$2$}}}),\\
 \label{2.1-1.2}
 [\widehat{\mathbf{\Uppsi}}^{\dagger}{\sml x_{\mbox{\tiny{$1$}}}\smr},\,\widehat{\mathbf{\Uppsi}}^{\dagger}{\sml x_{\mbox{\tiny{$2$}}}\smr}]&=&0,\\
 \label{2.1-1.3}
 [\widehat{\mathbf{\Uppsi}}{\sml x_{\mbox{\tiny{$1$}}}\smr},\,\widehat{\mathbf{\Uppsi}}{\sml x_{\mbox{\tiny{$2$}}}\smr}]&=&0,
\end{eqnarray}
for bosons, and the anti commutator algebra
\begin{eqnarray}
 \label{2.1-1.4}
 \{\widehat{\mathbf{\Uppsi}}{\sml x_{\mbox{\tiny{$1$}}}\smr},\,\widehat{\mathbf{\Uppsi}}^{\dagger}{\sml x_{\mbox{\tiny{$2$}}}\smr}\}&=&\delta(x_{\mbox{\tiny{$1$}}}-x_{\mbox{\tiny{$2$}}}),\\
 \label{2.1-1.5}
 \{\widehat{\mathbf{\Uppsi}}^{\dagger}{\sml x_{\mbox{\tiny{$1$}}}\smr},\,\widehat{\mathbf{\Uppsi}}^{\dagger}{\sml x_{\mbox{\tiny{$2$}}}\smr}\}&=&0,\\
 \label{2.1-1.6}
 \{\widehat{\mathbf{\Uppsi}}{\sml x_{\mbox{\tiny{$1$}}}\smr},\,\widehat{\mathbf{\Uppsi}}{\sml x_{\mbox{\tiny{$2$}}}\smr}\}&=&0,
\end{eqnarray}
for fermions.

Suppose that we want to populate the vertex at the position $x$ with $n$ particles. Mathematically we can write
\begin{equation}
 \label{2.1-3}
 {\lgl\widehat{\mathbf{\Uppsi}}^{\dagger}{\sml x\smr}\lgr}^{n}|\Omega\rangle=\sqrt{(n_x)!}|n_x\rangle,
\end{equation}
where $|n_{x}\rangle$ is the eigenstate of the operator $\hat{\bf n}_{x}=\widehat{\mathbf{\Uppsi}}^{\dagger}{\sml x\smr}\widehat{\mathbf{\Uppsi}}{\sml x\smr}$. For the moment we are only
interested by the number of particles at vertex $x$, any other property of particles such as modes are irrelevant.

Since we will shift particles between different vertices therefore in a discrete setting, the position of a vertex will be indexed ${x}_{\alpha}$ in order to distinguish different vertices. In that regards, when considering two vertices with different number of particles, we can write $|{n}_{x_{\alpha\mbox{\tiny{$1$}}}}\rangle$ and $|{n}_{x_{\alpha\mbox{\tiny{$2$}}}}\rangle$. For the sake of notations simplification we will use:
$|{n}_{\alpha_{\mbox{\tiny{$1$}}}}\rangle$ instead $|{n}_{x_{\alpha\mbox{\tiny{$1$}}}}\rangle$ and $|{n}_{\alpha_{\mbox{\tiny{$2$}}}}\rangle$ instead $|{n}_{x_{\alpha\mbox{\tiny{$2$}}}}\rangle$.

We consider $N$ indistinguishable bosons distributed on a $M$-vertices graph. The Hilbert space of such a system is spanned by 
\begin{equation}
\label{1}
D\sml \mbox{\small{$N,M$}}\smr=\bigg({\mbox{\small{$M+N-1$}}\atop{\mbox{\small{$N$}}}}\bigg)
\end{equation}
vectors $|{\bf n}\rangle=|n_1,\ldots, n_M\rangle=|{n}_{\alpha_{\mbox{\tiny{$1$}}}}\rangle\otimes|{n}_{\alpha_{\mbox{\tiny{$2$}}}}\rangle\otimes\ldots\otimes|{n}_{\alpha_{\mbox{\tiny{$M$}}}}\rangle$ of magnitude $N = \sum_{\alpha}^M n_{\alpha}$. In quantum optics and quantum matter, laser light have been used to confine cold atoms on lattices. In such a case the word configuration is used to indicate a specific way of arranging cold atoms on lattice \cite{Immanuel}. We use the same word configuration to indicate the  vectors $|{\bf n}\rangle$. The components of the $|{\bf n}\rangle$'s represent the position occupation numbers $n_{\alpha}$ of individual vertices.  The sum of the $n_{\alpha}$'s over the entire graph is equal to the total number of quantum walkers $N$. Individual possible configuration will be denoted as $|{\bf n}_{\ell}\rangle$, where $\ell=1,\ldots,D\sml \mbox{\small{$N,M$}}\smr$.

The set of all configuration is called Hilbert space of configurations.  In general, a state of a $M$-vertices graph 
containing $N$ particles can be represented as
\begin{equation}
 \label{2.2-1}
\big|\Psi\big\rangle=\sum_{\ell}C_{\ell}|{\bf n}_{\ell}\rangle,
\end{equation}
where $|{\bf n}_{\ell}\rangle$ generate a basis of the Hilbert space of many-particles on  a graph. We use Dirac bra ket notation for simplicity.
 In addition, we need an auxiliary Hilbert space (the so called coin's Hilbert space) spanned by the vectors specifying the directions of the edges connected to any given vertex.
 The number of edges ending at a vertex is called the vertex degree of this vertex. The degree of a graph is the largest of its vertices degrees \cite{Reinhard:01}. For a graph of degree $d$, the 
coin's Hilbert space is spanned by the basis $\{|v_{_1}\rangle,\ldots,|v_{d}\rangle\}$. These $d$ vectors are also known as 
chiralities \cite{Ambainis02} of the coin and we will elaborate more on them in the next section. The tensor product of the coin's Hilbert space and the configurations Hilbert space is equivalent to attaching a single coin to a configuration $|{\bf n}_{\ell}\rangle$. Therefore, a single coin is shared by all particles under the fixed configuration. The GMP states in Eq. (\ref{2.2-1}) augmented with the coins chiralities
 becomes
\begin{equation}
 \label{2.3}
|\Psi_{r}\rangle=\sum_{\ell}\sum_{i}\frac{C_{i\ell}^{r}}{\mathcal{K}_{r}}|v_{i},\,{\bf n}_{\ell}\rangle,
\end{equation}
where $|v_{i},\,{\bf n}_{\ell}\rangle=|v_{i}\rangle\otimes|{\bf n}_{\ell}\rangle$. 
The state $|\Psi_{\mbox{\tiny{r}}}\rangle$ represents the system of $N$ particles distributed on $M$ vertices of the graph, augmented with the directional amplitudes. We use the abbreviation 'GMP state' to indicate $|\Psi_{\mbox{\tiny{r}}}\rangle$. The index $r$ indicates the time step of the quantum walks and
$\mathcal{K}_{r}$ is the normalization constant 
\begin{equation}
 \label{2.4}
 [\mathcal{K}_{r}]^2=\sum_{\ell}\sum_{i}|C_{i\ell}^{r}|^{2}.
 \end{equation}
Since $|n_{\alpha}\rangle$ can be created from the vacuum $|\Omega\rangle$, see Eq. \eqref{2.1-3}, we have constructed the GMP state using field operator scheme \cite{Wigner:01}. 
Now we want to equip the vertices populations with some additional properties. To this end we recall that every particle
created at a given vertex is characterized by its mode ${\upeta}$. Let us consider a position field operator $\widehat{\mathbf{\Uppsi}}^{\dagger}{\sml {x}_{\alpha}\smr}$ and the related Fock creation operator ${\hat{\bf b}}_{\upeta}^{\dagger}$ as well as the conjugated operators $\widehat{\mathbf{\Uppsi}}{\sml {x}_{\alpha}\smr}$ and  ${\hat{\bf b}}_{\upeta}$:
\begin{equation}
 \label{2.5}
 \widehat{\mathbf{\Uppsi}}^{\dagger}{\sml {x}_{\alpha}\smr}=\sum_{\upeta}\frac{e^{-i{\varphi}_{\upeta}{x}_{\alpha}}}{\sqrt{N}} {\hat{\bf b}}_{\upeta}^{\dagger}\,\,\,{\mbox{and}}\,\,\,\widehat{\mathbf{\Uppsi}}{\sml {x}_{\alpha}\smr}=\sum_{\upeta}\frac{e^{i{\varphi}_{\upeta}{x}_{\alpha}}}{\sqrt{N}} {\hat{\bf b}}_{\upeta},
\end{equation} 
 where $\varphi_{\upeta}$ denotes the phase of a mode $\upeta$ and  ${x}_{\alpha}$ is the position. The boson Fock operators obey the following commutation rules
 \begin{equation}
 \label{eq:2.5.1}
  \big[{\hat{\bf b}}_{\upeta_{\mbox{\tiny{1}}}},\,{\hat{\bf b}}^{\dagger}_{\upeta_{\mbox{\tiny{2}}}}\big]=\delta_{{\upeta_{\mbox{\tiny{1}}}}{\upeta_{\mbox{\tiny{2}}}}},\,\,\,\big[{\hat{\bf b}}_{\upeta_{\mbox{\tiny{1}}}},\,{\hat{\bf b}}_{\upeta_{\mbox{\tiny{2}}}}\big]=0,\,\,\,\big[{\hat{\bf b}}^{\dagger}_{\upeta_{\mbox{\tiny{1}}}},\,{\hat{\bf b}}^{\dagger}_{\upeta_{\mbox{\tiny{2}}}}\big]=0.
 \end{equation} 
The boson Fock representation enables us to rewrite the GMP state in Eq.
(\ref{2.3}) in multimode boson Fock representation as:
\begin{equation}
 \label{2.6}
 |\Psi_{r}\rangle=\sum_{\ell}\sum_{j}\frac{C_{j\ell}^{r}}{\mathcal{K}_{r}}|v_{j}\rangle\bigotimes_{\alpha=1}^{M}{1\over{\sqrt{(n_{\alpha})!}}}\bigg[\sum_{\upeta}\frac{e^{-i{\varphi}_{\upeta}{x}_{\alpha}}}{\sqrt{N}} {\hat{\bf b}}_{\upeta}^{\dagger}\bigg]^{n_{\alpha}}|\Omega\rangle,   
 \end{equation}
 where $|\Omega\rangle$ is the vacuum state.
 %\vspace*{-0.75cm}
 %======================================
\section{The conditional shift operator}
\label{shift}
%=======================================
%\vspace*{-0.3cm}
Before entering in the design of our conditional shift operator we need to give some remarks about distinction between continuous time quantum walks and discrete time quantum walks. Continuous time quantum walks are analogous to classical diffusion process where the stochastic matrix is substituted by the adjacency matrix of the graph and the probabilities are replaced by the amplitudes. This makes the continuous quantum walks as the process described by the Schr\"{o}dinger equation where the Hamiltonian is the adjacency matrix \cite{Childs04}. Since the adjacency matrix is real and symmetric, the time evolution is unitary. Moreover it was proved that there is no discrete time process that is translationally invariant on a $d$-dimensional graph \cite{Meyer:01,Meyer:02}. Concerning our many-particle quantum walks, the conditional shift operator is the product of the field annihilation and creation operators, the adjacency matrix and the coin operator (Hadamard operator) as we will show in this section. During the implementation of the quantum walks, the action of the annihilation and creation operators in the conditional shifting of many-particles doesn't preserve the length of configurations and therefore destroys the unitarity. For this reason we need a step by step normalization of the GMP state. We also notice that the dimension of the Hilbert space of the discrete time quantum walks is greater than that of a continuous time quantum walks.

As it was already noted by several authors, the structure of the GMP state suggests that the operator implementation of the quantum walks 
is a two stages operation. The first stage is performed using the coin's tossing operation.
It is defined by a coin tossing operator that is unitary operator acting on the coin's Hilbert space. After the coin tossing the particle shifting is performed with respect to the selected direction. In a 
single particle quantum walk on a line, according to  \cite{Nayak:01, julia:01, julia:02}, the coin tossing is performed using the Hadamard gate 
\begin{equation}
\label{2.3-1}
{\bf H}={1\over{\sqrt{2}}}\left( \begin{array}{cc}
        1 & 1\\1 & -1\end{array} \right).
\end{equation}
On a line the coin Hilbert space is two dimensional (i.e. $d=2$) and spanned by vectors $|v_{1}\rangle$ and $|v_{2}\rangle$ that correspond to the left-hand side move and right-hand move respectively. Considering the vectors  $|v_{1}\rangle=\big({\mbox{\tiny{1}}\atop{\mbox{\tiny{0}}}}\big)$ and  $|v_{2}\rangle=\big({\mbox{\tiny{0}}\atop{\mbox{\tiny{1}}}}\big)$ of the coins Hilbert space, the coin tossing respectively transforms $|v_{1}\rangle$ and $|v_{2}\rangle$ in following ways
\begin{eqnarray}
 \label{2.3-3}
  {\bf H}|v_{1}\rangle&=&{1\over{\sqrt{2}}}(|v_{1}\rangle+|v_{2}\rangle),\\
  \label{2.3-4}
  {\bf H}|v_{2}\rangle&=&{1\over{\sqrt{2}}}(|v_{1}\rangle-|v_{2}\rangle).
\end{eqnarray}
This coin tossing operation is combined with the particle shifting operation where the shift is performed using the following operator
\begin{equation}
 \label{2.3-5}
  {\bf S}=|v_{2}\rangle\langle v_{2}|\otimes\sum_{\nu\in\mathbb{Z}}|\nu-1\rangle\langle\nu|+|v_{1}\rangle\langle v_{1}|\otimes\sum_{\nu\in\mathbb{Z}}|\nu+1\rangle\langle\nu|.
\end{equation}
This kind of shifting is well constructed for a single-particle quantum walk on the line where the structure of the 
graph is such that adjacent vertices are also successive vertices. In such a setup the projector
$|\nu+1\rangle\langle\nu|\equiv|\nu+1\rangle\langle0|\cdot|0\rangle\langle\nu|$
is equivalent to the product $\widehat{\mathbf{\Uppsi}}^{\dagger}{\sml {x}_{\nu+1}\smr}\widehat{\mathbf{\Uppsi}}{\sml {x}_{\nu}\smr}$ of field creation and annihilation operators. This line constructed scheme of coin tossing and particle shifting can be extended to other types of graphs and many-particle shifting. Bearing that in mind, we define the coin's tossing operator ${\bf H}_{d}$ as a $d$-order Hadamard operator given by the following formula
\begin{equation}
 \label{2.3-6}
  {\bf H}_{d}=\sum_{j,k}h_{j k}|v_{j}\rangle\langle v_{k}|\,\,\,\,\,j,\,k\,\in\,1,2,3,\ldots,d.
\end{equation}
These $h_{\jmath k}$ are normalized roots of the unity of order $d$. 
The operation on the coin's Hilbert space provides the direction to the quantum walk. In addition, we need the shifting operation which implements the
movement of quantum walkers from theirs previous positions (one configuration) to their next positions (other configuration) at each step. Such an operation over the graph uses the property of 
graph connectivity and the field creation and annihilation operations. In fact, particles can only be shifted between adjacent vertices, therefore 
the adjacency matrix is involved in the shifting operation. For the $M$-vertices graph, the adjacency matrix 
is a $M$-dimension square matrix with entries
\begin{eqnarray}
A_{\mu\nu} &=&  \begin{cases}
1  &   ~~  \text{when $\mu$ and $\nu$ are connected,} \\
0  &   ~~ \text{when $\mu$ and $\nu$ are not connected.} 
\end{cases}\nonumber
\end{eqnarray}
In addition to the adjacency matrix, the field creation and annihilation 
operators interfere every time when a particle is shifted from vertex $\mu$ to vertex $\nu$. In other words, the annihilation operator acts on vertex
$\mu$ removing one occupant while the creation operator acts on vertex $\nu$ adding one more occupant. 
For example for the cyclic graph in Fig.\ref{fig:123}a, the adjacency matrix is a $10\times10$ matrix ${\bf A} = \big(A_{\nu\mu}\big)$; $\mu,\,\nu\,=\,1,\ldots,10$ of the form
\begin{equation}
 \label{4-2.1.2}
{\bf A}=\left(
\begin{array}{ccccccc}
 0 	& 1 	& \cdots & 0 	& 1 	\\
 1	& 0 	& \cdots & 0 	& 0 	\\
 \vdots &\vdots &\ddots &\vdots &\vdots \\
 0 	& 0 	& \cdots	& 0 	& 1 	\\
 1 	& 0 	& \cdots	& 1 	& 0
\end{array}
\right).
\vspace*{-0.2cm}
\end{equation}
Graphs considered in quantum walks are undirected and quantum particles on them can spread in all directions. For the cyclic graph in Fig.\ref{fig:123}a. the quantum walks on each edge $(\mu,\,\nu)$ are realized both ways clockwise and counterclockwise and this property enables to write its
adjacency matrix as the sum of its two mutually transposed components  
\begin{equation}
 \label{2.3-10}
  {\bf A}={\bf A}_{\mbox{\tiny{L}}}+{\bf A}_{\mbox{\tiny{L}}}^{\mbox{\tiny{T}}},
\end{equation}
where
\begin{equation}
\label{2.3-10.1}
{\bf A}_{\mbox{\tiny{L}}}=\left(
\begin{array}{ccccccc}
 0 	& 0 	& \cdots & 0 	& 1 	\\
 1	& 0 	& \cdots & 0 	& 0 	\\
 \vdots &\vdots &\ddots &\vdots &\vdots \\
 0 	& 0 	& \cdots	& 0 	& 0 	\\
 0 	& 0 	& \cdots	& 1 	& 0
\end{array}
\right).
\vspace*{-0.2cm}
\end{equation}
${\bf A}_{\mbox{\tiny{L}}}$ is the adjacency matrix for directed cyclic graph corresponding to counterclockwise walks on this graph. 
Applying this property of the adjacency matrix to the vertex particle annihilation and creation operations enables to write the shift operator for the
cyclic graph in Fig.\ref{fig:123}a in the following form
\begin{equation}
\label{2.3-14}
{\bf S}=\mathcal{C}^{\dagger}{\bf A}_{\mbox{\tiny{L}}}\mathcal{C}\otimes|v_{1}\rangle\langle v_{1}|+\mathcal{C}^{\dagger}{\bf A}_{\mbox{\tiny{L}}}^{\mbox{\tiny{T}}}\mathcal{C}\otimes|v_{2}\rangle\langle v_{2}|,
\end{equation}
where:
\begin{equation}
\label{2.3-15}
\begin{cases}
\mathcal{C} &=\big(\widehat{\mathbf{\Uppsi}}{\sml {x}_{\mbox{\tiny{$1$}}}\smr},\ldots,\widehat{\mathbf{\Uppsi}}{\sml {x}_{\mbox{\tiny{$M$}}}\smr}\big)^{\mbox{\tiny{T}}}, \\
\mathcal{C}^{\dagger} & =\big(\widehat{\mathbf{\Uppsi}}^{\dagger}{\sml {x}_{\mbox{\tiny{$1$}}}\smr},\ldots,\widehat{\mathbf{\Uppsi}}^{\dagger}{\sml {x}_{\mbox{\tiny{$M$}}}\smr}\big).  
\end{cases}
\end{equation}
The similar decomposition of the adjacency matrix as in Eq. \eqref{2.3-10} is valid for the line graph. For two degrees undirected graphs, two directions of quantum walks correspond to two components of the adjacency matrix ${\bf A}$ in Eq. \eqref{4-2.1.2}. In general let us consider a $d$ degrees graph, each direction induces an automorphism of the graph. The automorphism of graph are elements of the permutation group defined  on the graph. On any given vertex $\mu$ we define an permutation in direction $k$, $P^k\sml \mu\smr\equiv\nu$. The set of all permutations $P^k\sml \mu\smr\equiv\nu$ over the entire graph defines a class of equivalence in the permutation group. Such a class of equivalence is a graph. The adjacency matrix $(A_{\nu\mu}^{k})$ of such a graph corresponding to direction $k$ is a component of the adjacency matrix of the graph $(A_{\nu\mu})$. For example for the Petersen graph given in Fig.\ref{fig:123}c one can define four directions i.e. $d = 4$. The first direction $\mu\rightarrow\nu$ if $\mu-\nu=-1$ and the second direction $\mu\rightarrow\nu$ if $\mu-\nu=+1$.  The third direction $\mu\rightarrow\nu$ if $\mu-\nu=-5$, the fourth direction $\mu\rightarrow\nu$ if $\mu-\nu=+5$. The undirected move on each direction generates two mutually transposed components of the adjacency matrix. Thus for the Petersen graph we can have four components ${\bf A}^{k}=(A_{\nu\mu}^{k})$  such that: ${\bf A}^{1}=({\bf A}^{2})^{\mbox{\tiny{T}}}$ and ${\bf A}^{3}=({\bf A}^{4})^{\mbox{\tiny{T}}}$.  In general the action of the permutation group over the graph splits the adjacency matrix into $d$ components. 
 \begin{figure*}[t]
  \centering
  \begin{subfigure}[b]{0.28\textwidth}
    \centering
    \includegraphics[height = 1.0\textwidth,width=1.1\textwidth]{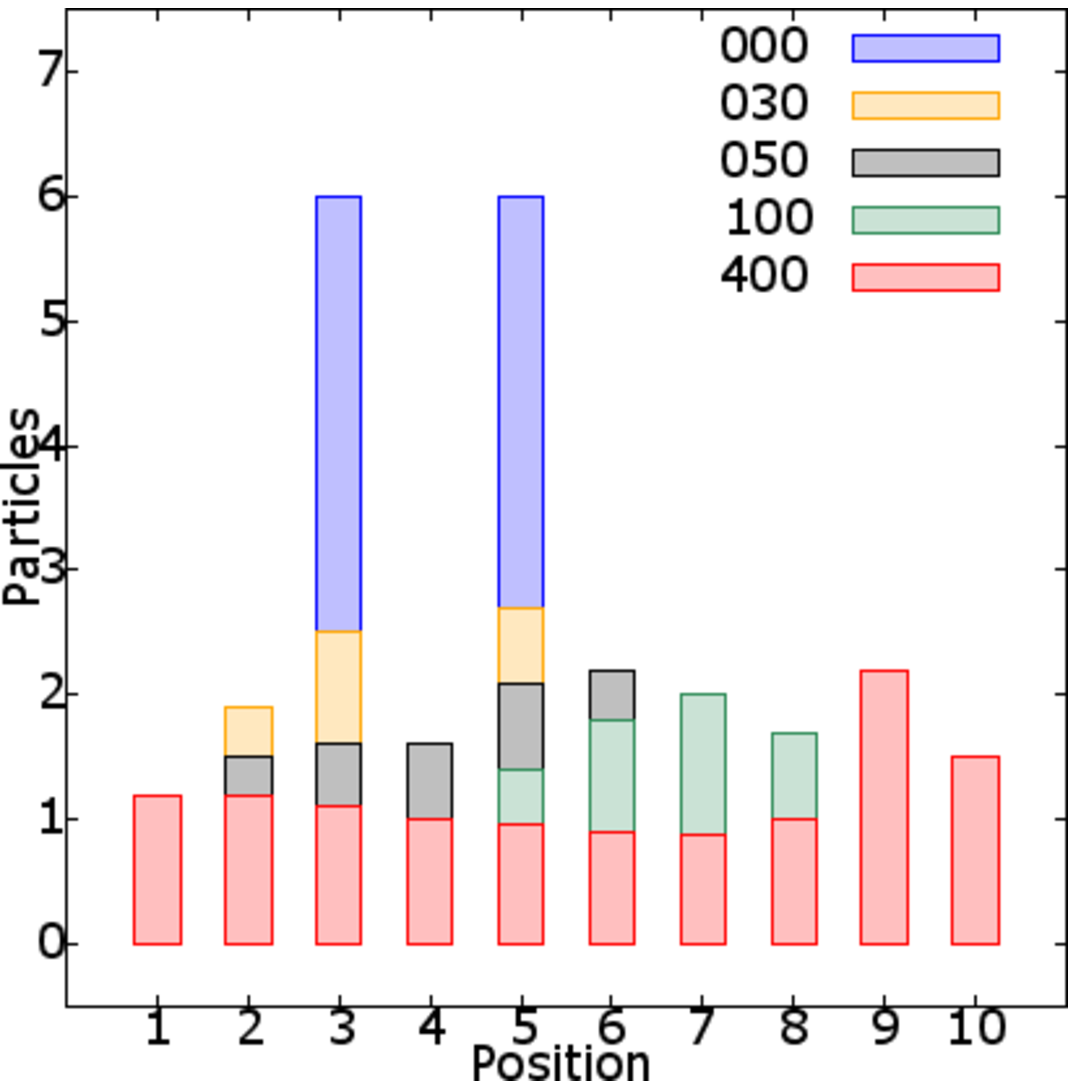}
    \caption{density: cycle\label{fig:4}}
  \end{subfigure}
  \hspace*{0.8cm}
  \begin{subfigure}[b]{0.28\textwidth}
    \centering
    \includegraphics[height = 1.0\textwidth,width=1.1\textwidth]{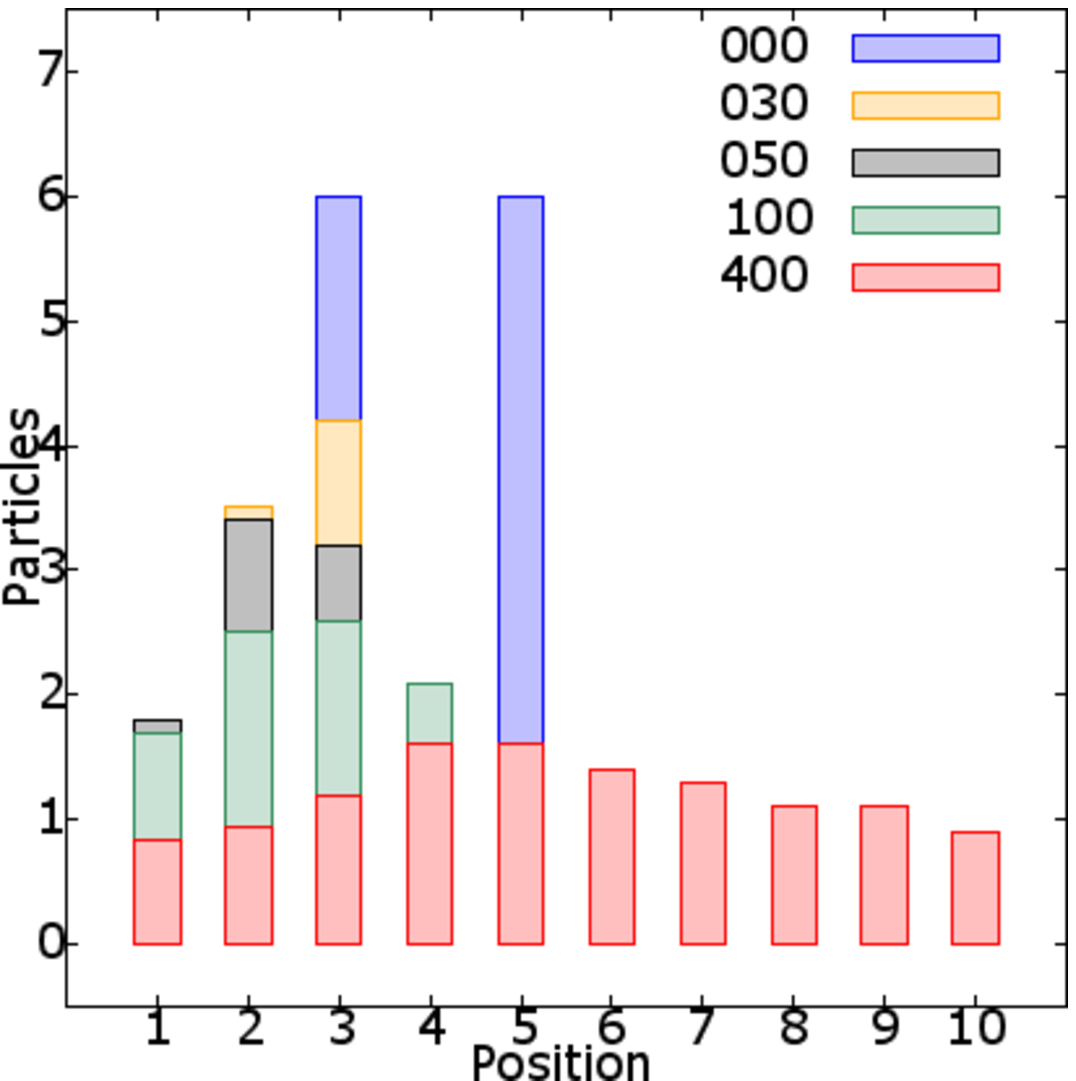}
    \caption{density: double hexagon\label{fig:5}}
  \end{subfigure}
   \hspace*{0.8cm}
  \begin{subfigure}[b]{0.28\textwidth}
    \centering
    \includegraphics[height = 1.0\textwidth,width=1.1\textwidth]{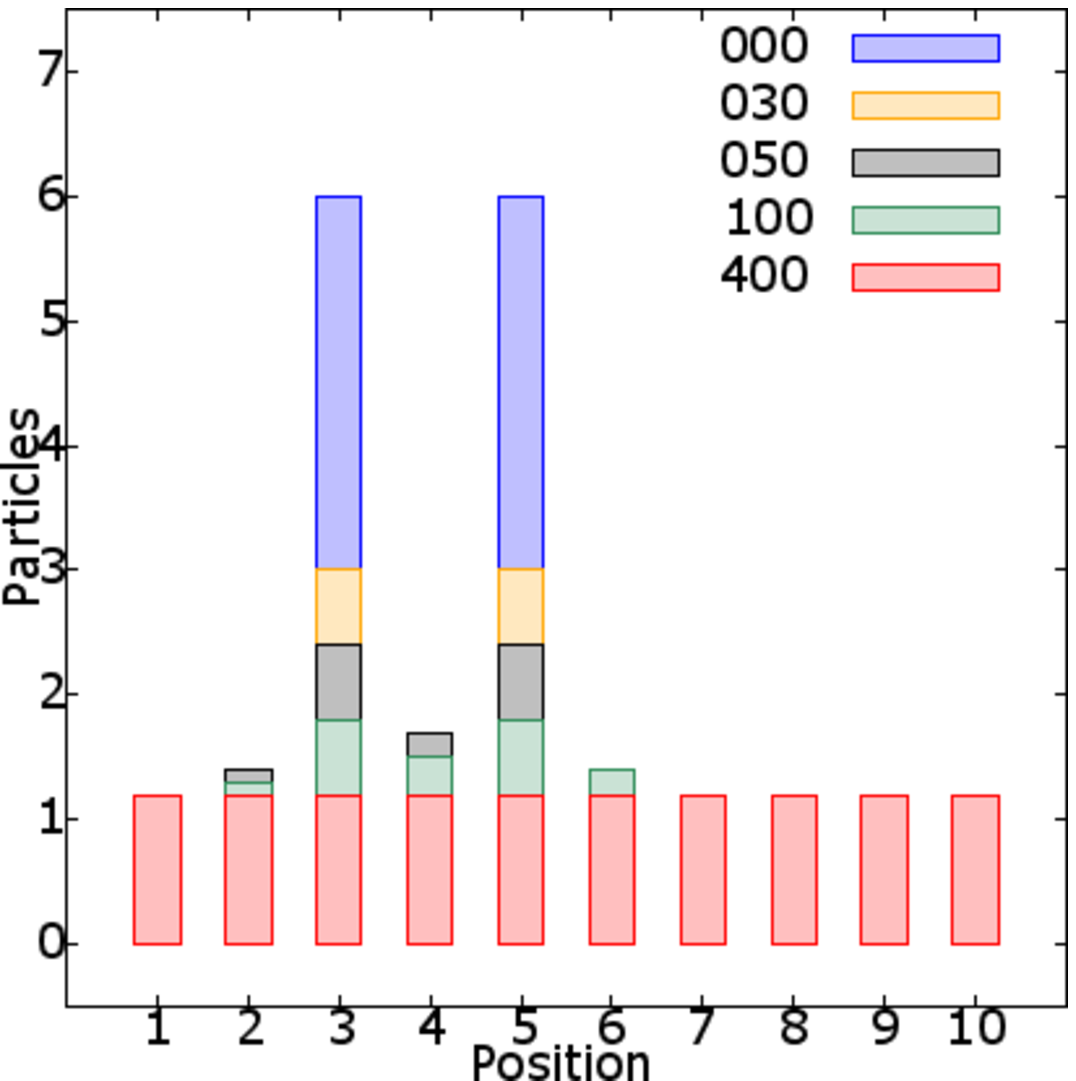}
    \caption{density: Petersen\label{fig:6}}
  \end{subfigure}
  \begin{subfigure}[b]{0.29\textwidth}
    \centering
    \includegraphics[height = 1.0\textwidth,width=1.1\textwidth]{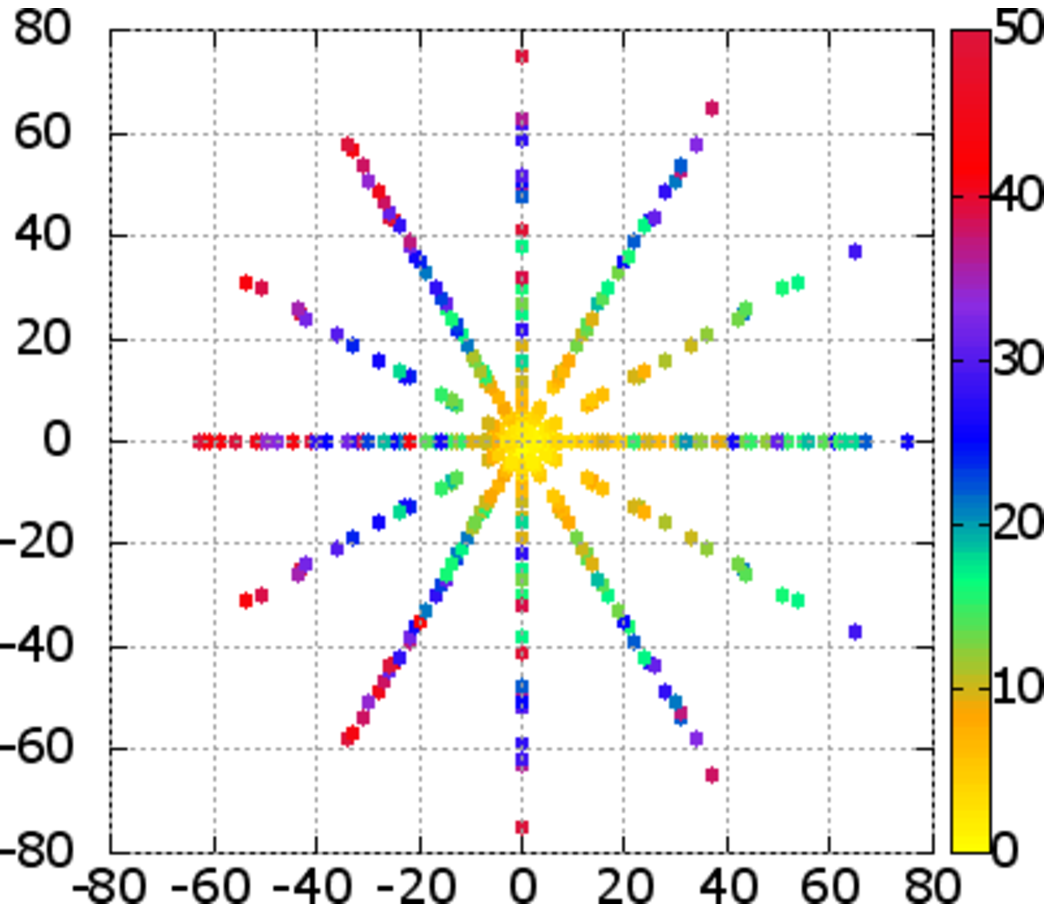}
    \caption{12 modes phase space walks\label{fig:7}}
  \end{subfigure}
  \hspace*{0.6cm}
  \begin{subfigure}[b]{0.29\textwidth}
    \centering
    \includegraphics[height = 1.0\textwidth,width=1.1\textwidth]{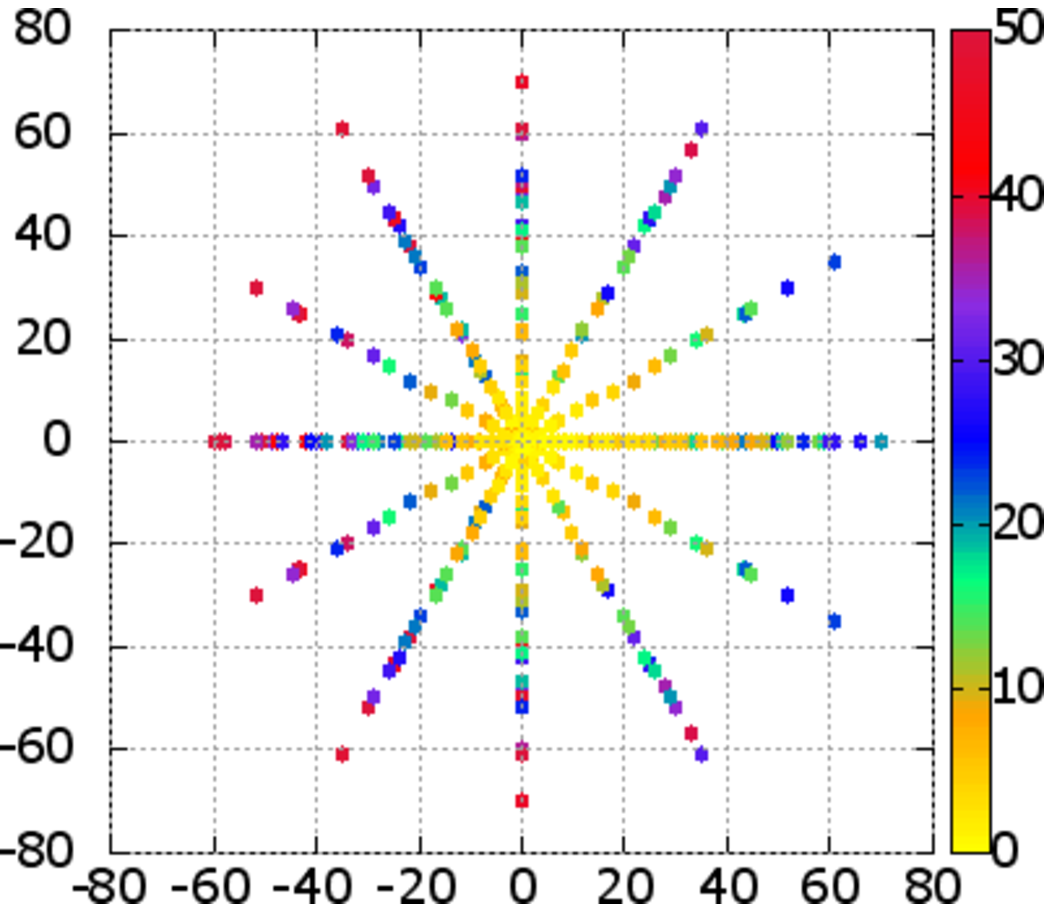}
    \caption{12 modes phase space walks\label{fig:8}}
  \end{subfigure}
   \hspace*{0.6cm}
  \begin{subfigure}[b]{0.29\textwidth}
    \centering
    \includegraphics[height = 1.0\textwidth,width=1.1\textwidth]{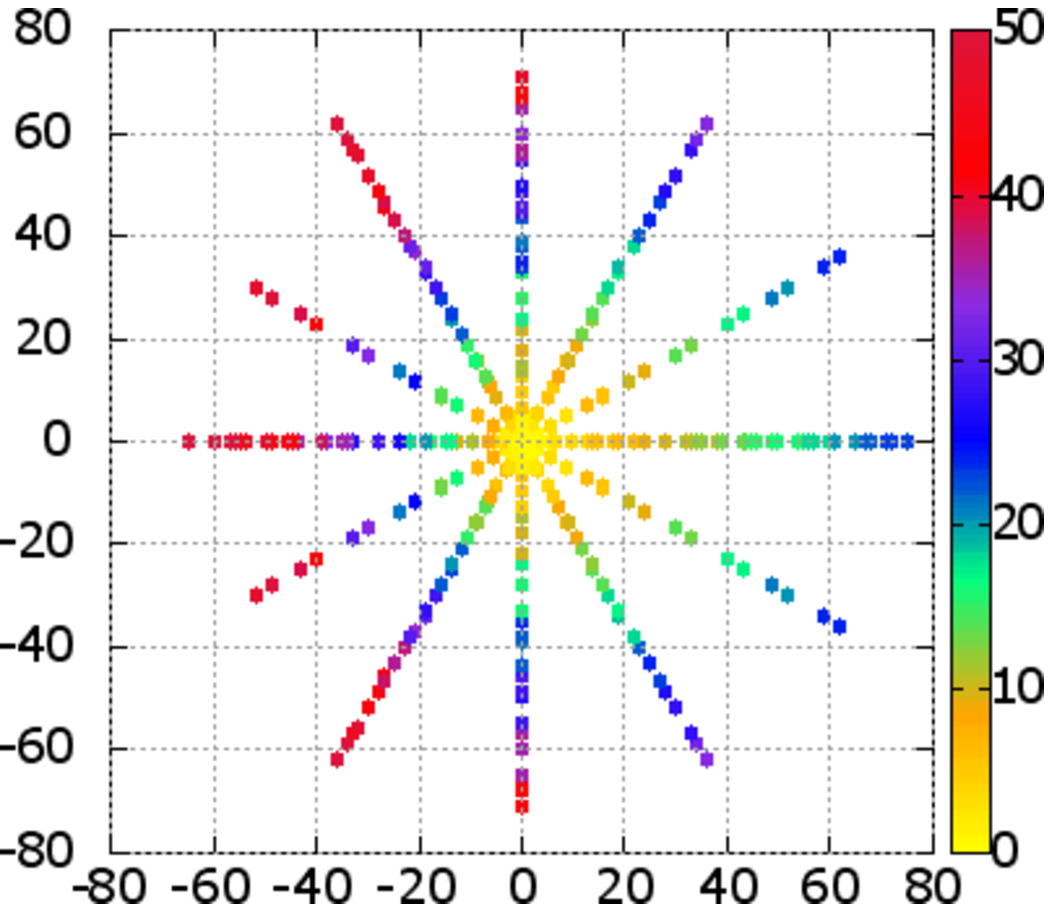}
    \caption{12 modes phase space walks\label{fig:9}}
  \end{subfigure}
  \caption{Selected steps average particle numbers (density) on vertices and phase space evolution during 400 steps of 12 quantum walkers on 10 vertices graphs:  (a), (d) -- cyclic, (b), (e) -- double hexagon and (c), (f) -- Petersen. The phase space walks spread from an initial single point at the center and split into modes during evolution. We have: position $\langle x_{\upeta}\sml r\smr\rangle$ on the x-axis,  momentum $\langle p_{\upeta}\sml r\smr\rangle$ on y-axis and energy $\langle E_{\upeta}\sml r\smr\rangle$ is on the colors bar. }
\vspace*{-0.2cm}
\end{figure*}
Therefore we generalize the shift operator given in Eq.  \eqref{2.3-14} to the following form
\begin{equation}
 \label{2.3-17}
  {\bf S}=\sum_{\mu,\nu}\sum_{k}\widehat{\mathbf{\Uppsi}}^{\dagger}{\sml {x}_{\nu}\smr}A_{\nu\mu}^{k}\widehat{\mathbf{\Uppsi}}{\sml {x}_{\mu}\smr}\otimes|v_k\rangle\langle v_k|.
\end{equation}
Here $A_{\nu\mu}^{k}$ indicates the action of the permutation group of the graph on its adjacency matrix. Such an operation associates an element $A_{\nu\mu}$ with the chirality $k$ of the coins. Consequently action of permutation group operation coordinates the choice of vertices on which the boson 
annihilation and boson creation act. On the line or the cyclic graph in Fig.\ref{fig:123}a, the permutation becomes just the transposition. The product of the Hadamard operator ${\bf H}_{d}$ defined in Eq. \eqref{2.3-6} and the shift operator ${\bf S}$ in Eq.\eqref{2.3-17}, gives the  conditional shift operator $\widehat{\mathcal{S}}$
\begin{equation}
 \label{2.3-18}
  \widehat{\mathcal{S}}=\sum_{\mu,\nu}\sum_{k}\widehat{\mathbf{\Uppsi}}^{\dagger}{\sml {x}_{\nu}\smr}A_{\nu\mu}^{k}\widehat{\mathbf{\Uppsi}}{\sml {x}_{\mu}\smr}\otimes\sum_{j}h_{jk}|v_j\rangle\langle v_k|.
\end{equation}
In the case of the cyclic graph shown in Fig.\ref{fig:123}a, we have $d=2,\,\,M=10,\,\,h_{11}=h_{12}=h_{12}={1\over{\sqrt{2}}}$ and $h_{22}=-{1\over{\sqrt{2}}}$. In such a case the matrix transposition 
operation is the only symmetry splitting the adjacency matrix therefore ${\bf A}_{\mbox{\tiny{L}}}=\big(A_{\nu\mu}^{\mbox{\tiny{1}}}\big)$ and ${\bf A}_{\mbox{\tiny{L}}}^{\mbox{\tiny{T}}}=\big(A_{\nu\mu}^{\mbox{\tiny{2}}}\big)$ with $\mu,\nu=1,\ldots,10$.
It follows
\begin{equation}
 \label{2.3-19}
  \widehat{\mathcal{S}}=\sum_{\mu,\nu=1}^{10}\sum_{k=1}^{2}\widehat{\mathbf{\Uppsi}}^{\dagger}{\sml {x}_{\nu}\smr}A_{\nu\mu}^{k}\widehat{\mathbf{\Uppsi}}{\sml {x}_{\mu}\smr}\otimes\sum_{j=1}^{2}h_{jk}|v_j\rangle\langle v_k|.
\end{equation}
Suppose that at a $r^{\mbox{\tiny{th}}}$ step such a system is represented by a GMP state in Eq.\eqref{2.6}. Then one step later it becomes
 \begin{equation}
 \label{3-12.2}
 |\Psi_{r+{\mbox{\tiny{1}}}}\rangle=\widehat{\mathcal{S}}|\Psi_{r}\rangle=\sum_{\ell_{\mbox{\tiny{1}}}}\sum_{j}\frac{C_{j\ell_{\mbox{\tiny{1}}}}^{r+1}}{\mathcal{K}_{r+1}}\big|v_{j}{\bf n}_{\ell_{\mbox{\tiny{1}}}}\big\rangle,
 \end{equation}
 where
 \begin{eqnarray}
 \label{3-13}
  C_{j\ell_{\mbox{\tiny{1}}}}^{r+1}&=&\sum_{\mu,\nu}\sum_{k}\frac{h_{jk}C_{k\ell}^{r}}{\mathcal{K}_{r}}\mbox{\large{$f$}}_{\mu\nu}^{k},\\
 \label{3-14}
 |{\bf n}_{\ell_{\mbox{\tiny{1}}}}\rangle&\sim&{\bf S}\,|{\bf n}_{\ell}\rangle,\\
 \label{3-15}
 \mbox{\large{$f$}}_{\mu\nu}^{k}&=&A_{\nu\mu}^{k}h_{jk}\sqrt{{n}_{\mu}({n}_{\nu}+1)},\\
 \label{3-16}
 [\mathcal{K}_{r+1}]^{2}&=&\sum_{\ell_{\mbox{\tiny{1}}}}\sum_{j}|C_{j\ell_{\mbox{\tiny{1}}}}^{r+1}|^{2}.
\end{eqnarray}
The term $\mbox{\large{$f$}}_{\mu\nu}^{k}$ contains residuals of field operators actions and depends on graph's  adjacency matrix. It shows how the graph structure is encoded in the GMP state
during the evolution. In addition, at every step the GMP state must be normalized that is why the normalization constant $\mathcal{K}_{r}$ is step $r$ dependent. Considering the GMP state obtained in Eq.\eqref{3-12.2} we observe that the conditional shifting plays the role in redefining the amplitude of the configuration over the GMP state. The triple sum in the recursion relations in Eq.\eqref{3-13} defines the relation between the amplitudes $C_{j\ell_{\mbox{\tiny{1}}}}^{r+1}$ and $C_{k\ell}^{r}$ during the step implementation and couples all the parent configurations ${\bf n}_{\ell}$ involved in the induction of the new configurations ${\bf n}_{\ell_{\mbox{\tiny{1}}}}$ that constitute the evolved GMP state.  In other words, the whole dynamics of many-particle quantum walks becomes walks over the configurations Hilbert space.  
 \begin{figure*}[t]
  \centering
  \begin{subfigure}[b]{0.30\textwidth}
    \centering
    \includegraphics[height = 0.8\textwidth,width=1.13\textwidth]{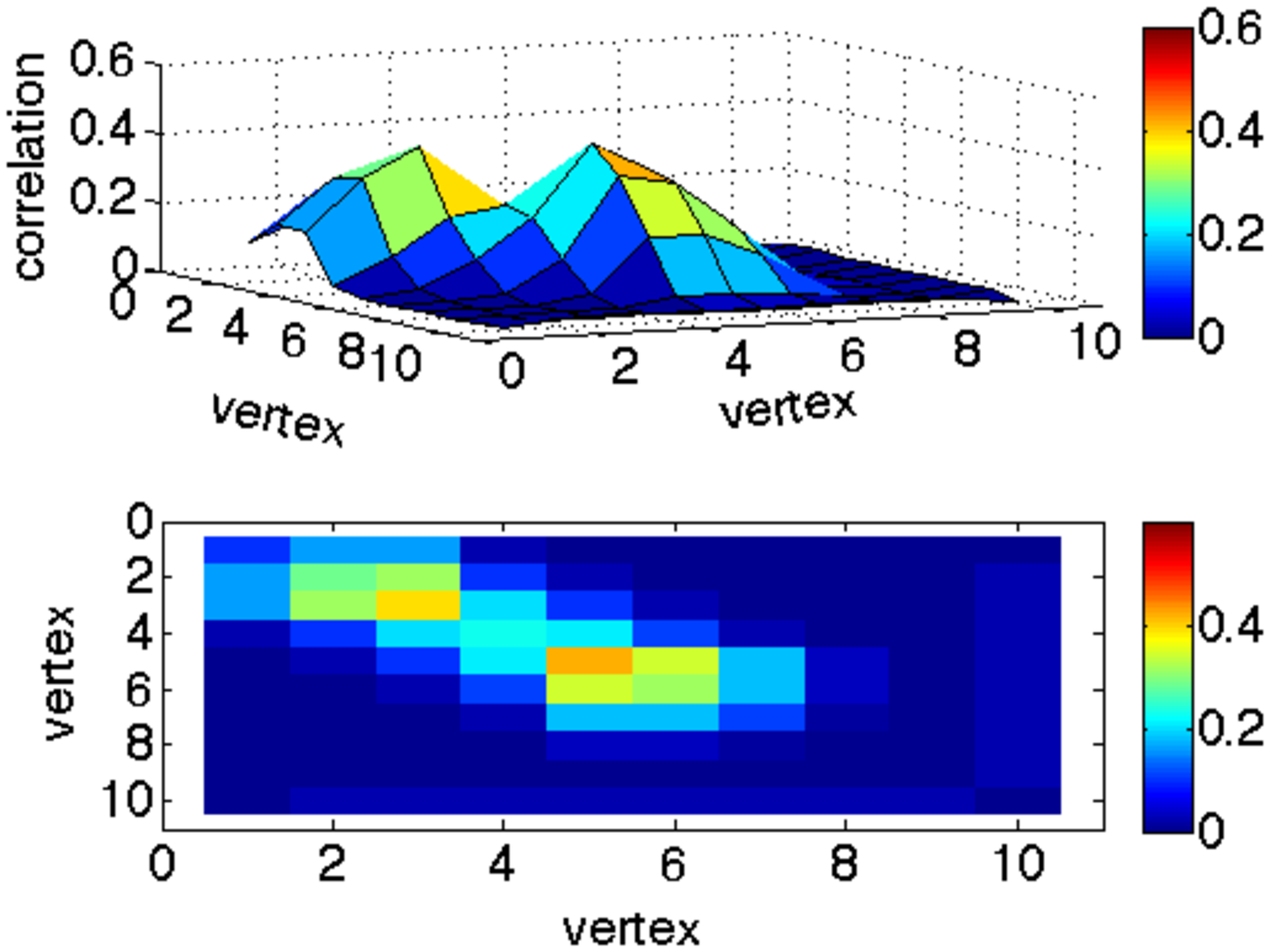}
    \caption{$30^{\mbox{\tiny{th}}}$ step; cyclic graph\label{fig:11}}
  \end{subfigure}
  \hspace*{0.1cm}
  \begin{subfigure}[b]{0.30\textwidth}
    \centering
    \includegraphics[height = 0.8\textwidth,width=1.13\textwidth]{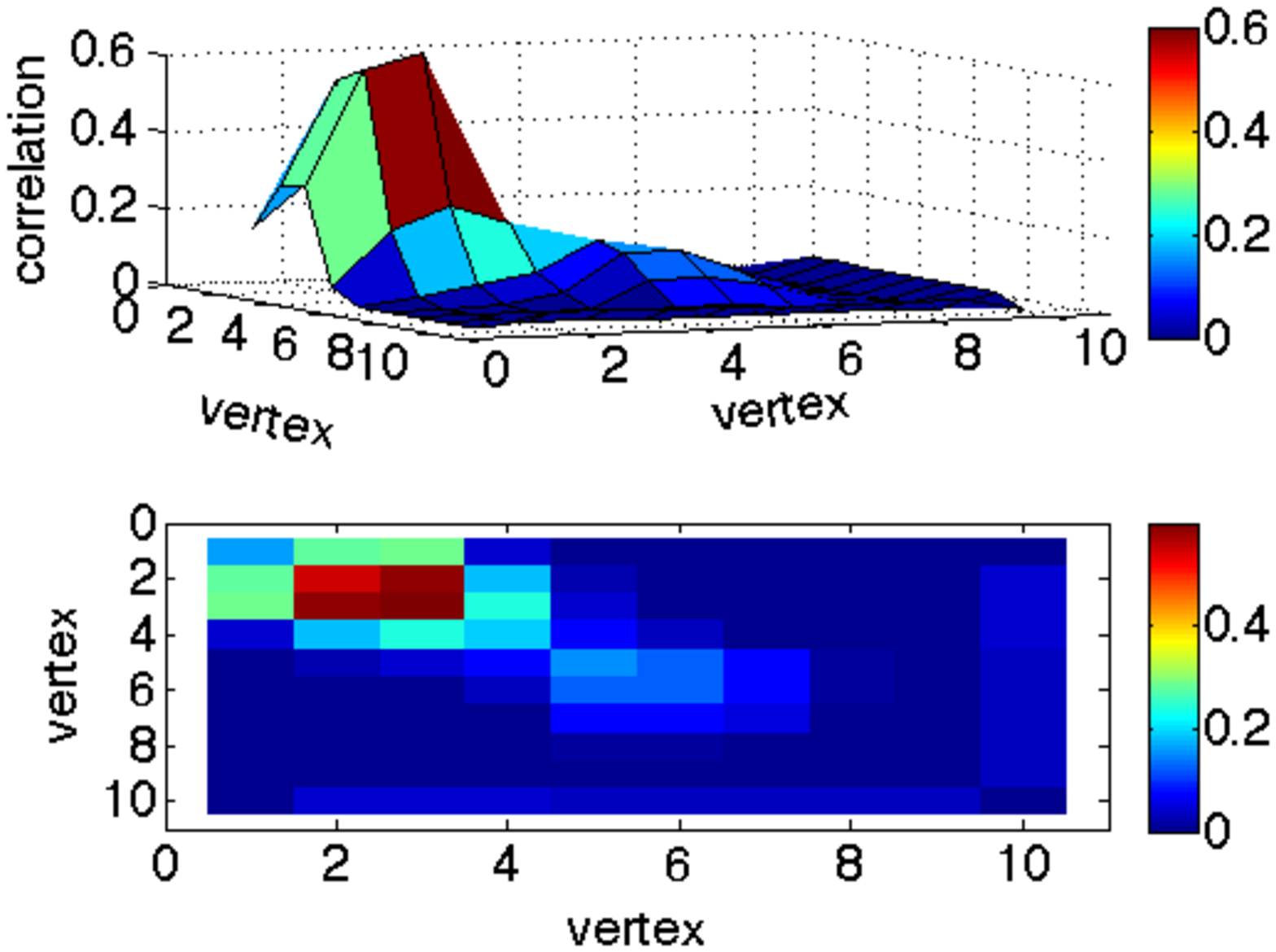}
    \caption{$30^{\mbox{\tiny{th}}}$ step; double hexagon graph\label{fig:12}}
  \end{subfigure}
   \hspace*{0.1cm}
  \begin{subfigure}[b]{0.30\textwidth}
    \centering
    \includegraphics[height = 0.8\textwidth,width=1.13\textwidth]{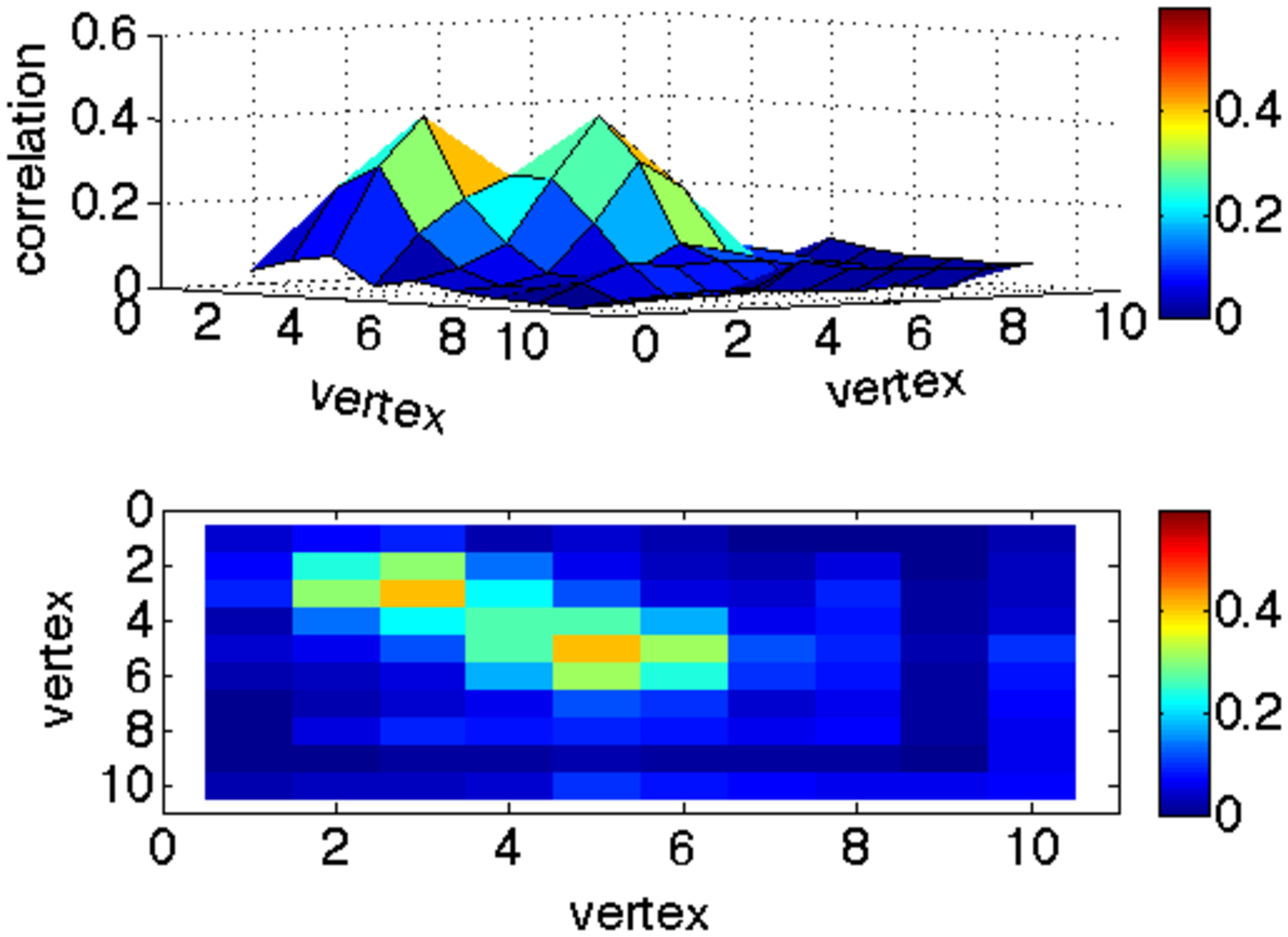}
    \caption{$30^{\mbox{\tiny{th}}}$ step; Petersen graph\label{fig:13}}
  \end{subfigure}
  \begin{subfigure}[b]{0.30\textwidth}
    \centering
    \includegraphics[height = 0.8\textwidth,width=1.13\textwidth]{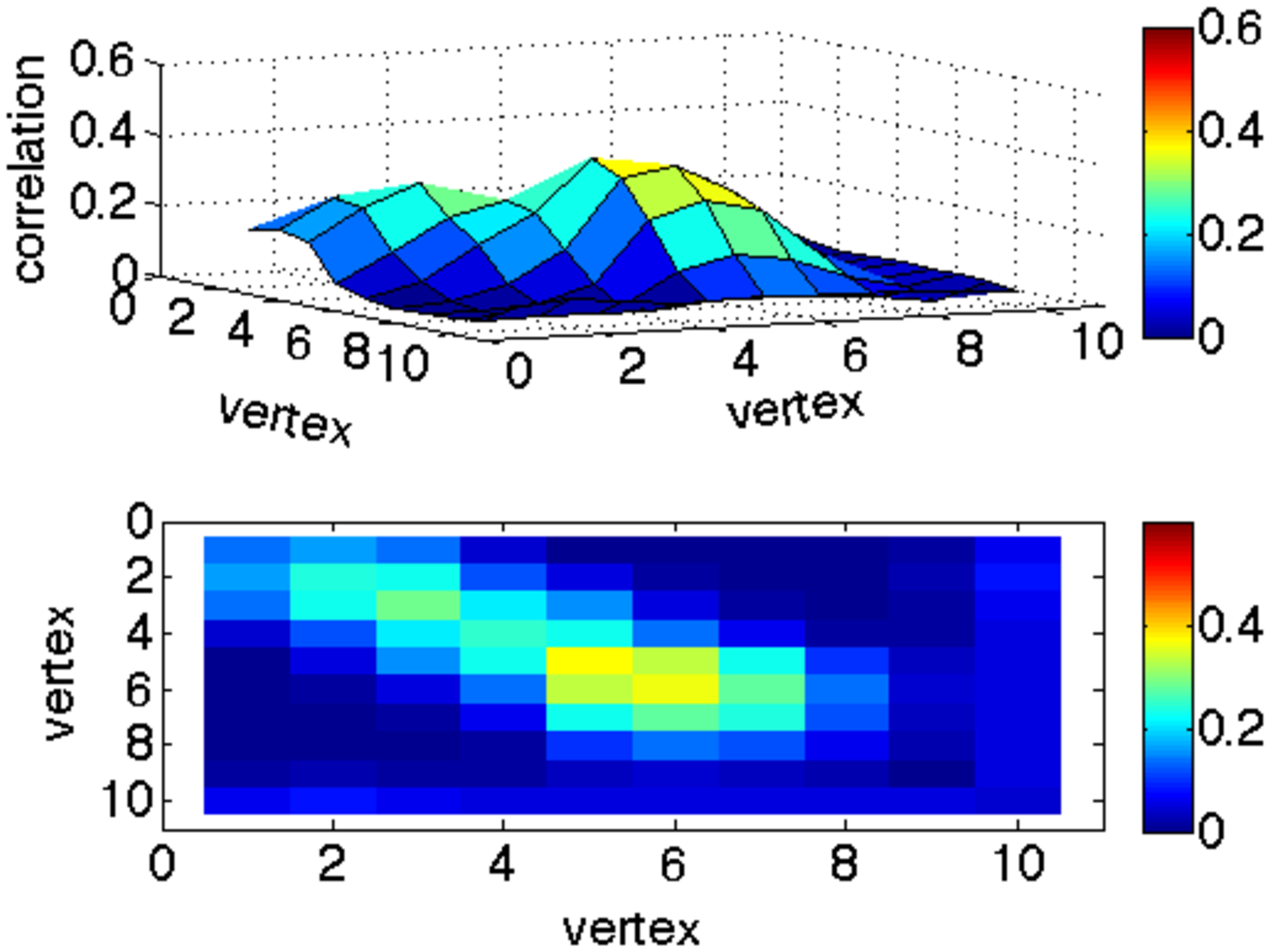}
    \caption{$50^{\mbox{\tiny{th}}}$ step; cyclic graph\label{fig:21}}
  \end{subfigure}
  \hspace*{0.1cm}
  \begin{subfigure}[b]{0.30\textwidth}
    \centering
    \includegraphics[height = 0.8\textwidth,width=1.13\textwidth]{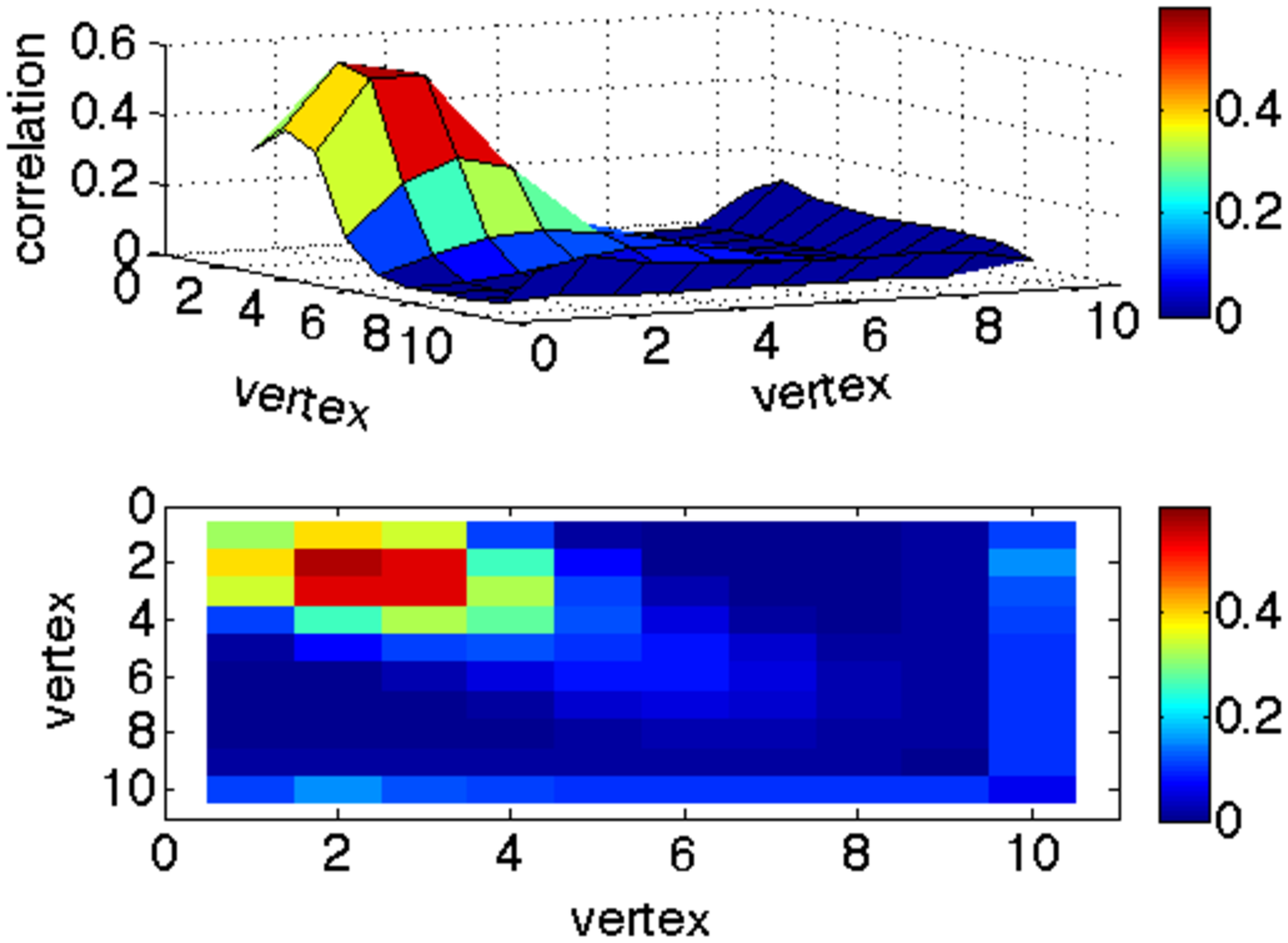}
    \caption{$50^{\mbox{\tiny{th}}}$ step; double hexagon graph\label{fig:22}}
  \end{subfigure}
   \hspace*{0.1cm}
  \begin{subfigure}[b]{0.30\textwidth}
    \centering
    \includegraphics[height = 0.8\textwidth,width=1.13\textwidth]{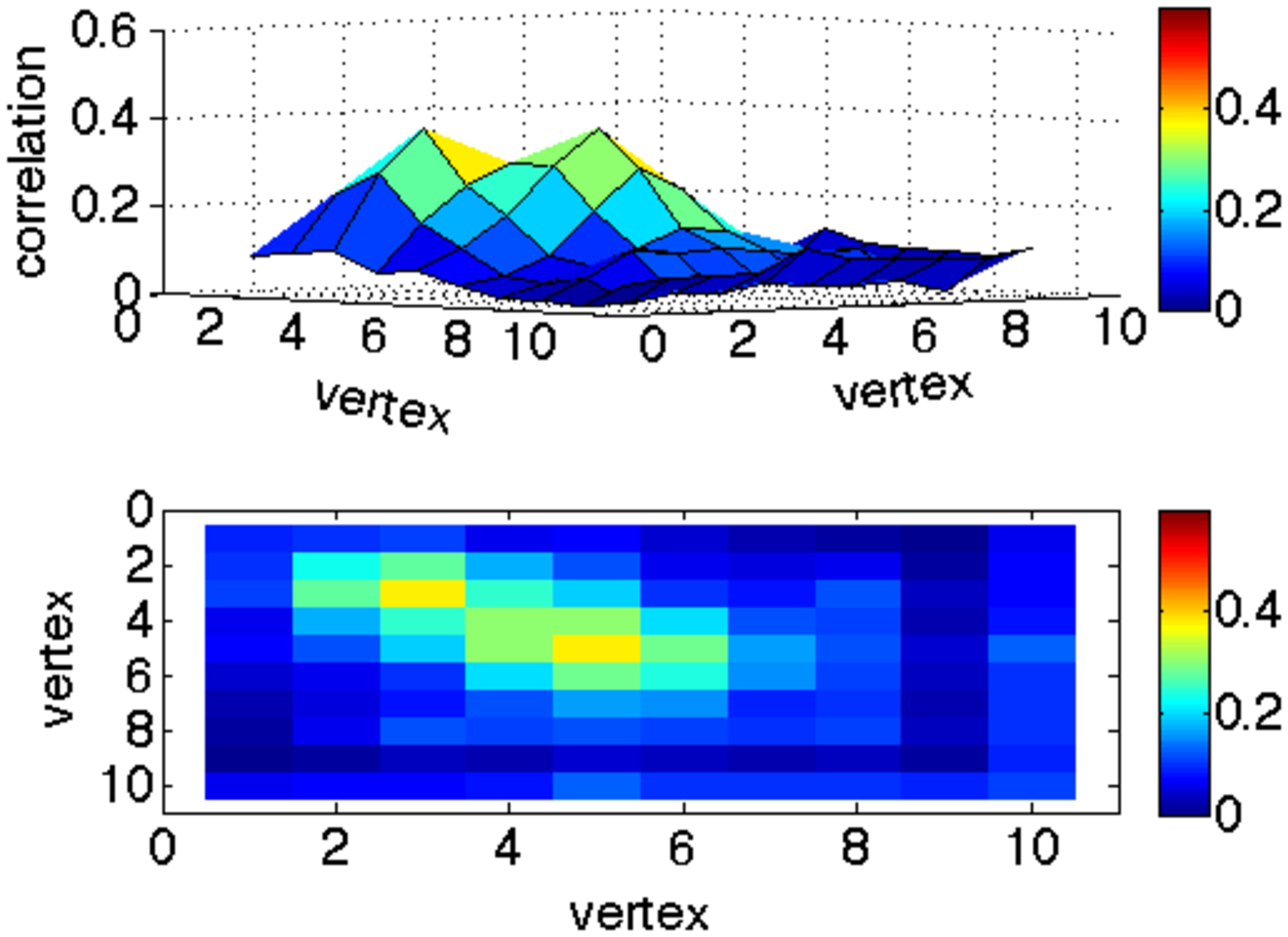}
    \caption{$50^{\mbox{\tiny{th}}}$ step; Petersen graph\label{fig:23}}
  \end{subfigure}
  \begin{subfigure}[b]{0.30\textwidth}
    \centering
    \includegraphics[height = 0.8\textwidth,width=1.13\textwidth]{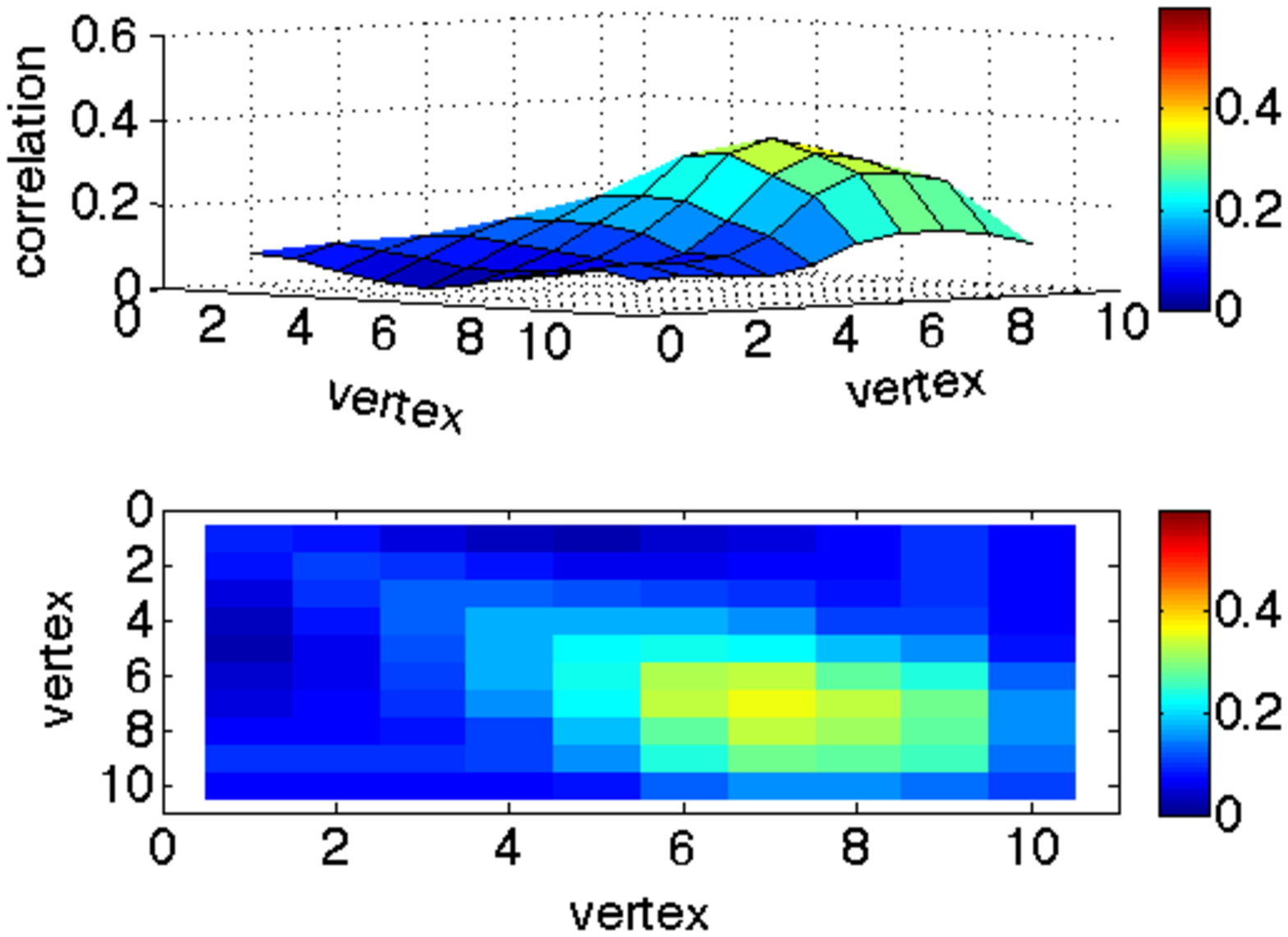}
    \caption{$100^{\mbox{\tiny{th}}}$ step; cyclic graph\label{fig:31}}
  \end{subfigure}
  \hspace*{0.1cm}
  \begin{subfigure}[b]{0.30\textwidth}
    \centering
    \includegraphics[height = 0.8\textwidth,width=1.13\textwidth]{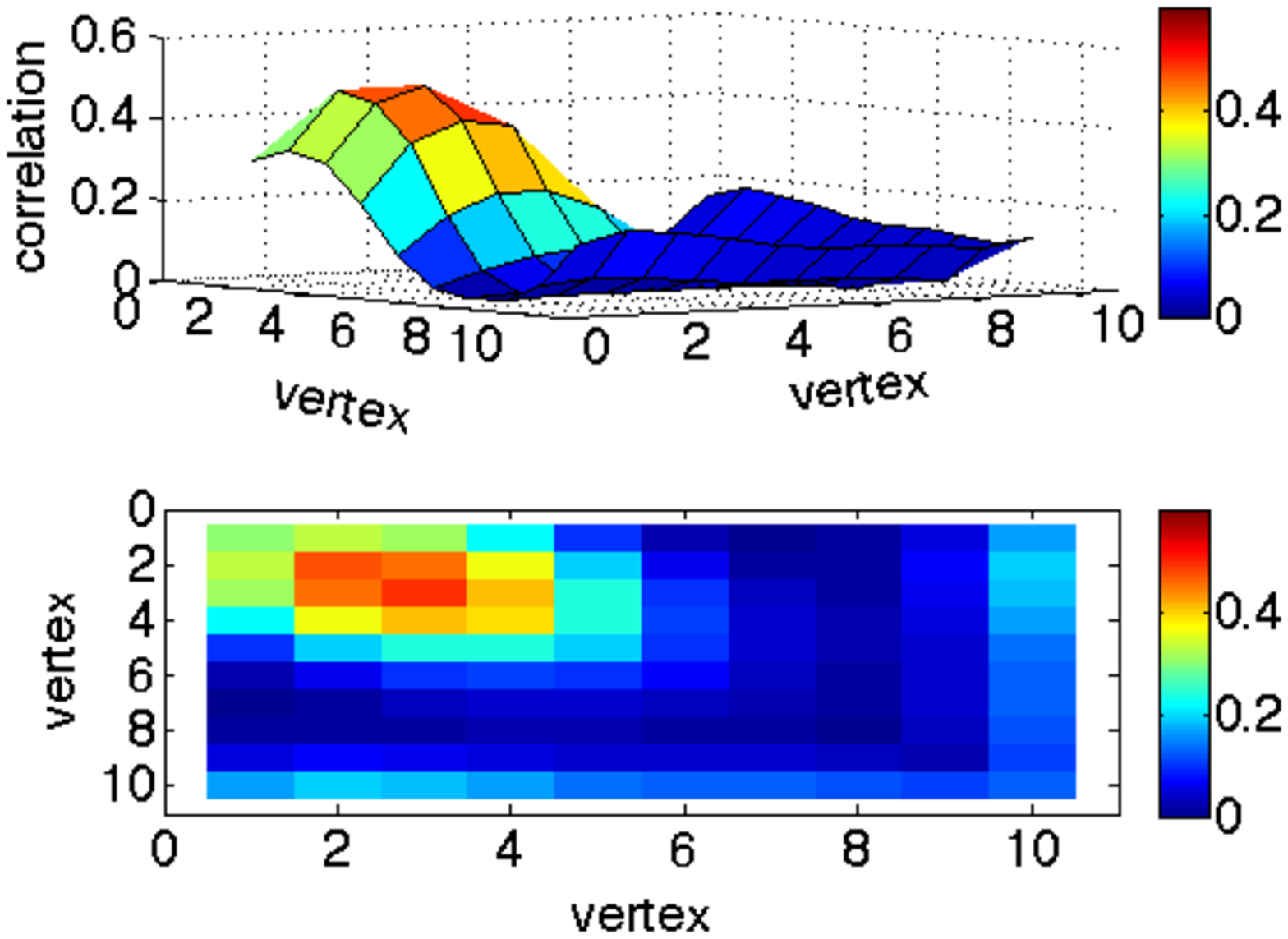}
    \caption{$100^{\mbox{\tiny{th}}}$ step; double hexagon graph\label{fig:32}}
  \end{subfigure}
   \hspace*{0.1cm}
  \begin{subfigure}[b]{0.30\textwidth}
    \centering
    \includegraphics[height = 0.8\textwidth,width=1.13\textwidth]{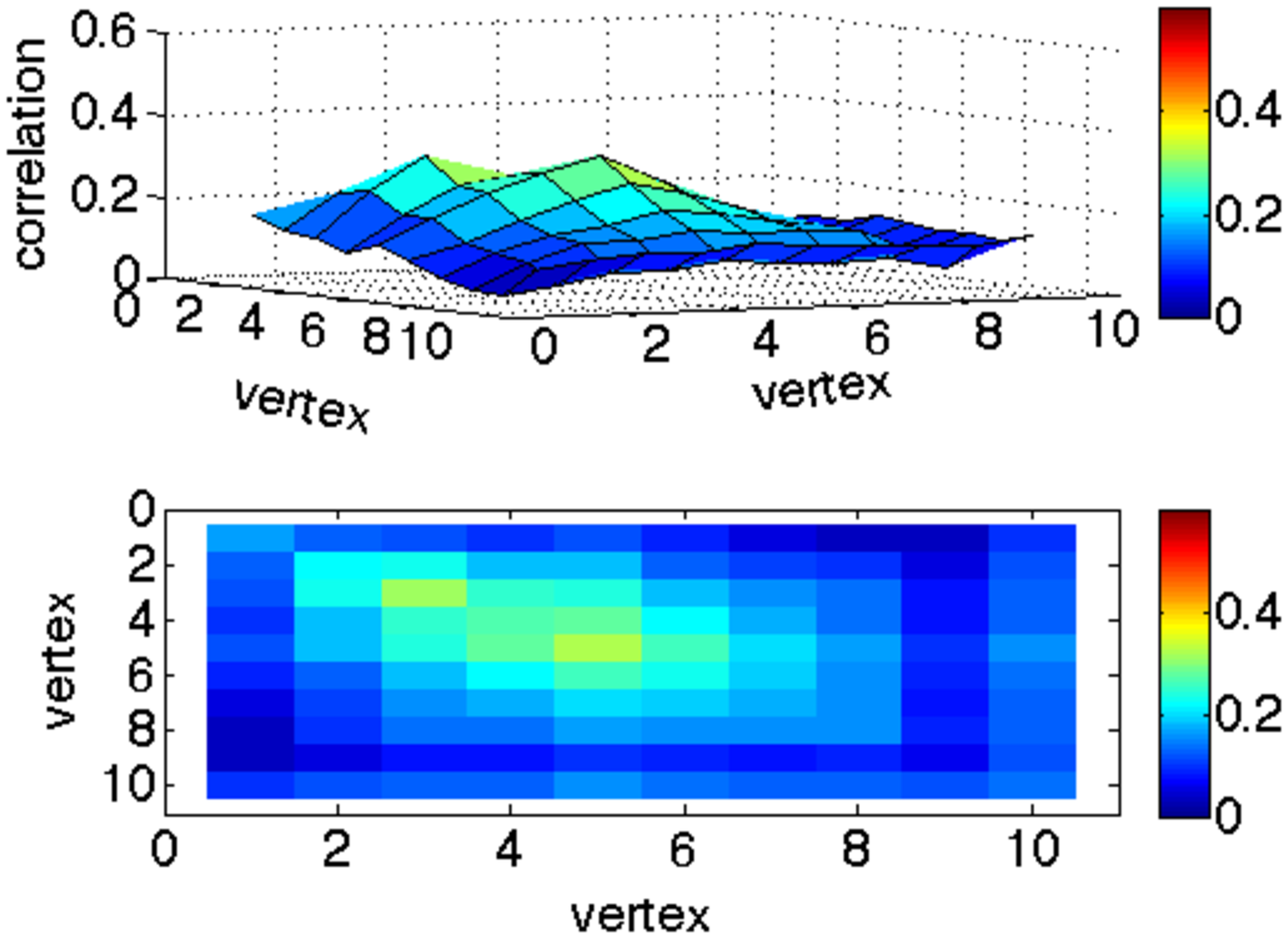}
    \caption{$100^{\mbox{\tiny{th}}}$ step; Petersen graph\label{fig:33}}
  \end{subfigure}
  \begin{subfigure}[b]{0.30\textwidth}
    \centering
    \includegraphics[height = 0.8\textwidth,width=1.13\textwidth]{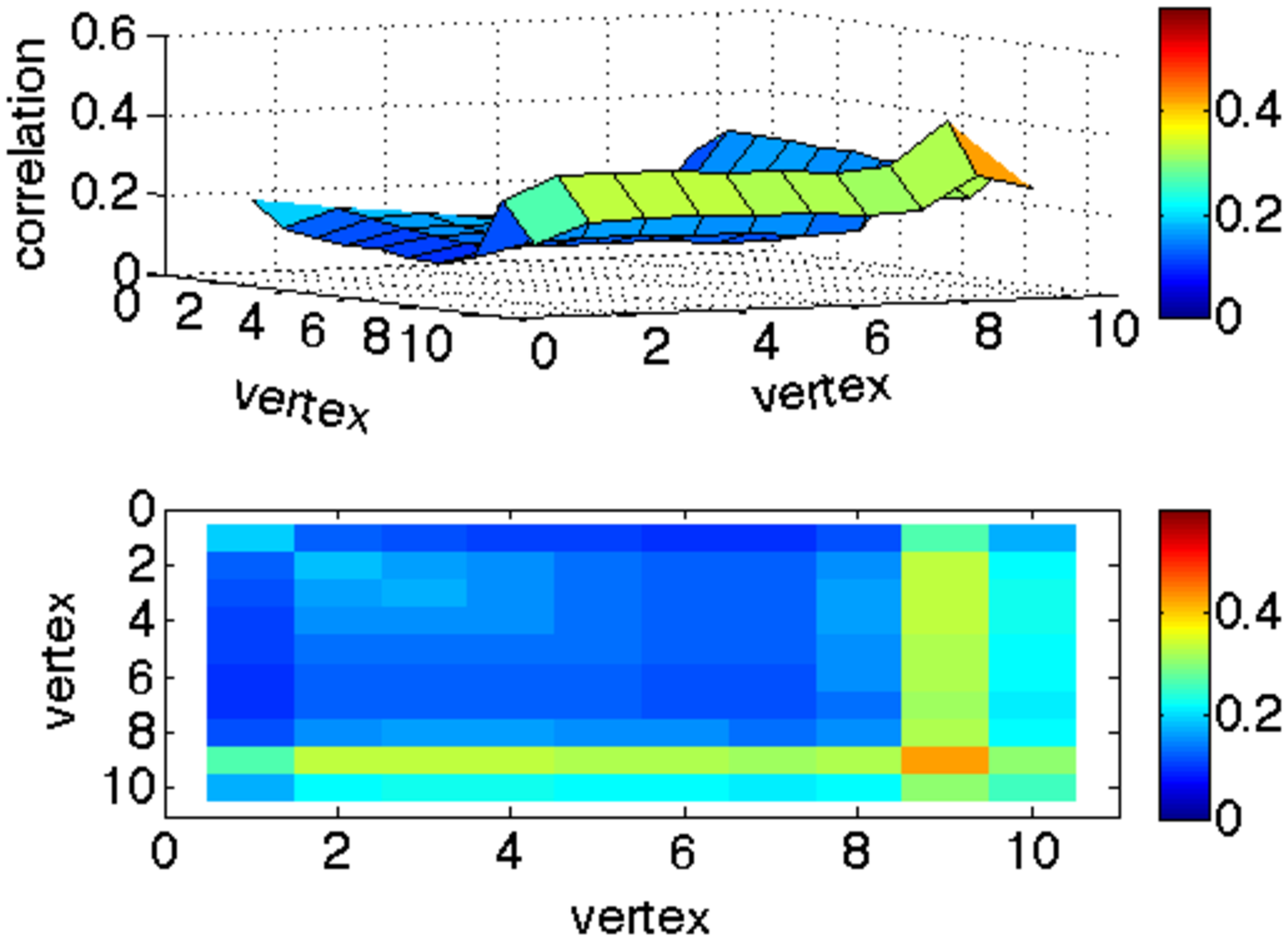}
    \caption{$400^{\mbox{\tiny{th}}}$ step; cyclic graph\label{fig:41}}
  \end{subfigure}
  \hspace*{0.1cm}
  \begin{subfigure}[b]{0.30\textwidth}
    \centering
    \includegraphics[height = 0.8\textwidth,width=1.13\textwidth]{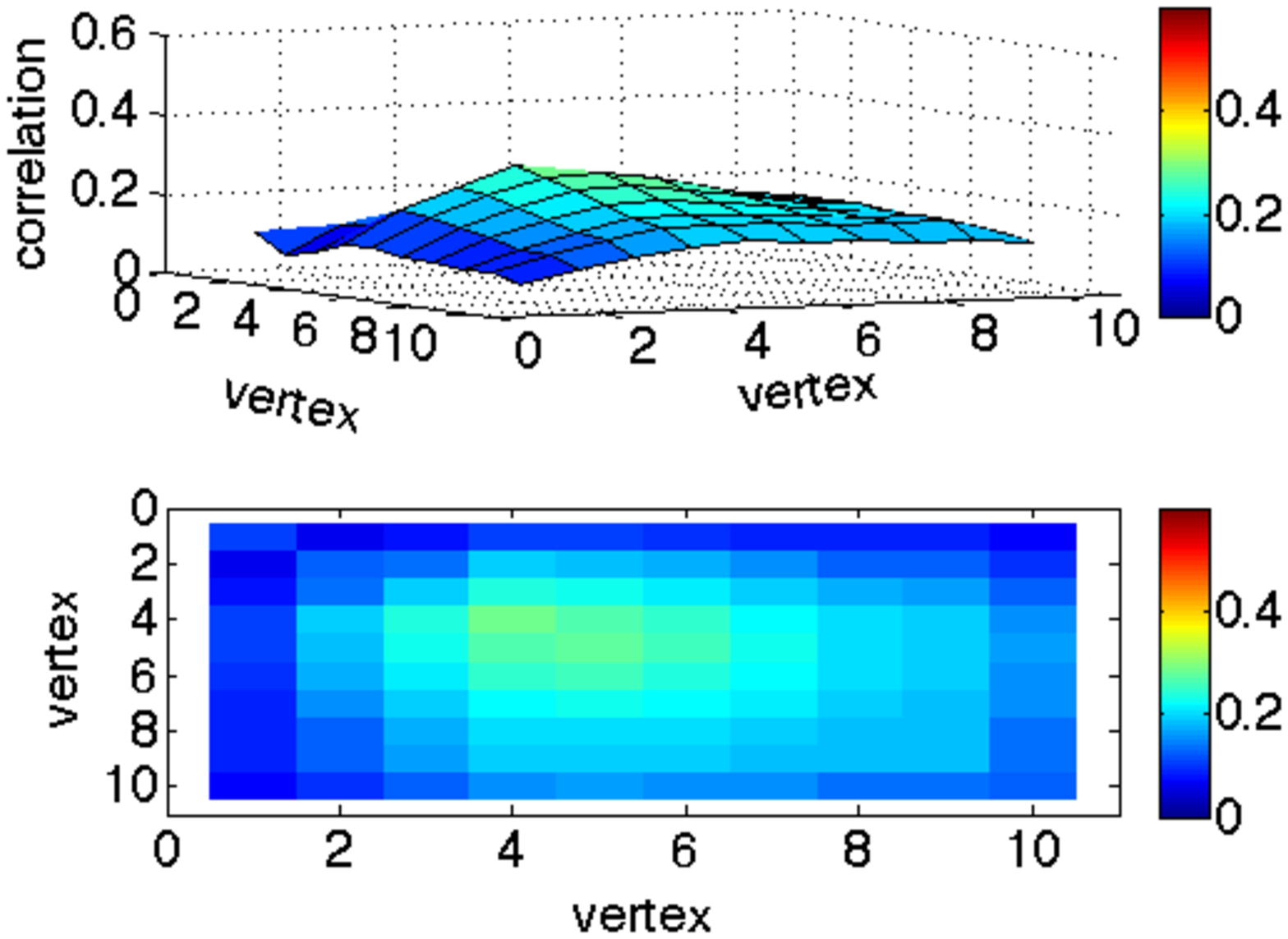}
    \caption{$400^{\mbox{\tiny{th}}}$ step; double hexagon graph\label{fig:42}}
  \end{subfigure}
   \hspace*{0.1cm}
  \begin{subfigure}[b]{0.30\textwidth}
    \centering
    \includegraphics[height = 0.8\textwidth,width=1.13\textwidth]{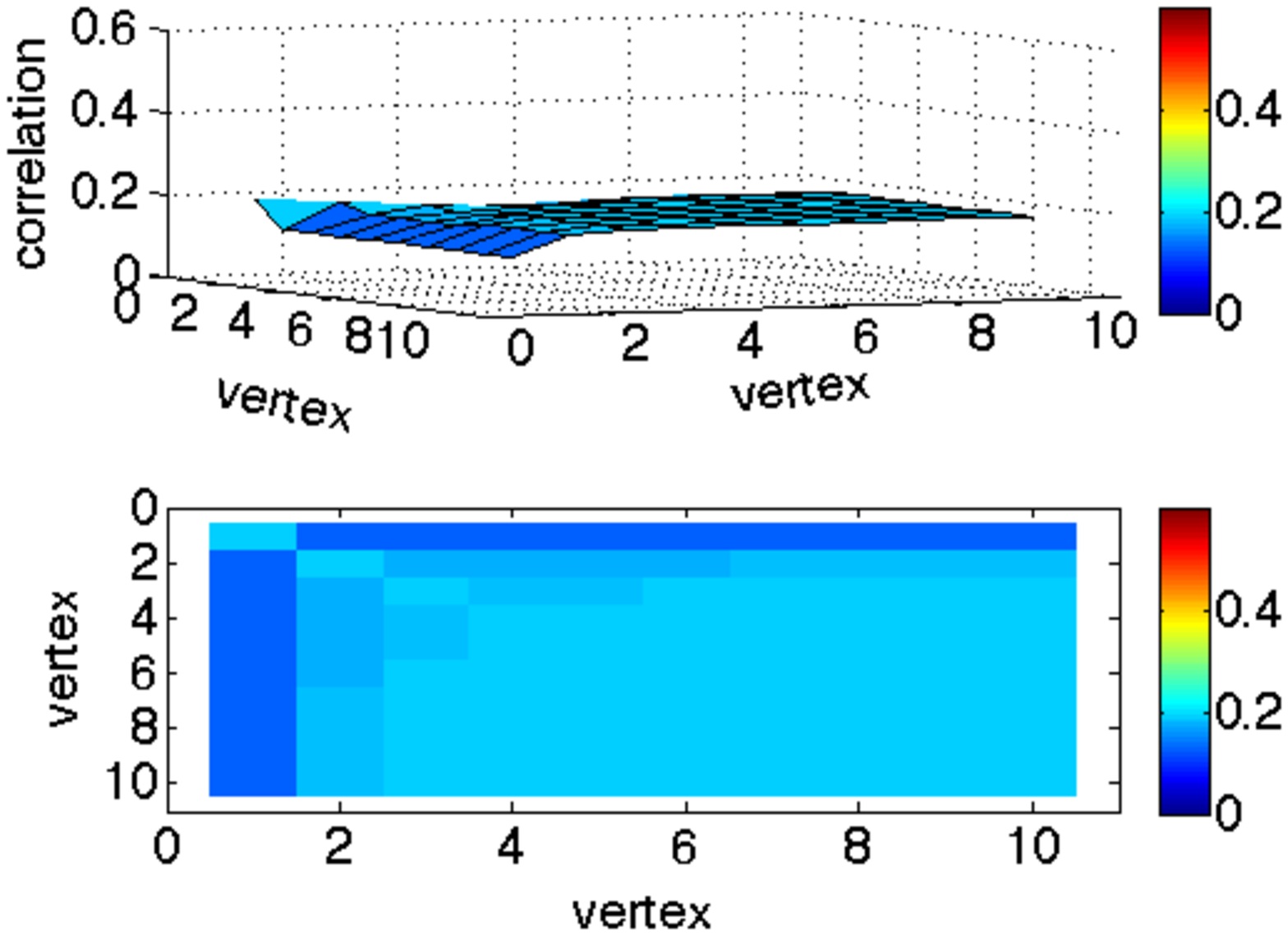}
    \caption{$400^{\mbox{\tiny{th}}}$ step; Petersen graph\label{fig:43}}
  \end{subfigure}
  \caption{Second order vertex-vertex correlations defined in Eq. \eqref{3-34} obtained during simulations of 12 quantum walkers at the $\mbox{30}^{\mbox{\tiny{th}}}$ step, $\mbox{50}^{\mbox{\tiny{th}}}$ step, $\mbox{100}^{\mbox{\tiny{th}}}$ step and $\mbox{400}^{\mbox{\tiny{th}}}$ step: for the cyclic graphs (a)--(d)--(g)--(j),  for the double hexagon graphs (b)--(d)--(h)--(k)  and for the Petersen graph (c)--(e)--(i)--(l). }
\vspace*{-0.2cm}
\end{figure*} 
In addition, mathematical expression of the evolution of amplitudes can be used as indicators of interfering
parent configurations dragging the system in the unexplored regions of the configurations Hilbert space or trapping the system in a specific subspace of this Hilbert space.   
%\vspace*{-0.75cm}
%=============================
\section{Results and conclusions}
%==============================
The probability of reaching a specific configuration ${\bf n}_{\ell}$ at the step $r$ is given by
\begin{equation}
 \label{3-27}
  P_{\ell}^{r}=\sum_{j}\bigg|\frac{C_{j\ell}^{r}}{\mathcal{K}_{r}}\bigg|^{2}.
\end{equation}
It is also the joint probability of the $M$-tuples occupation numbers $n_{\alpha}$'s over the graph.
The $q^{\mbox{\tiny{th}}}$ moment of the occupation number on a vertex $\alpha$ is given by: 
\begin{equation}
\label{3-28}
 \langle (n_{\alpha}\sml r\smr)^{q}\rangle=\sum_{\ell}(n_{\alpha})^{q}P_{\ell}^{r},
\end{equation} 
where $ \langle \cdot\rangle$ denote the mean value.
These moments are computationally accessible for any value of $q$. Frequently only $q=1,2$ and $3$ are used. The first moment  represents the vertex expected particles occupation numbers $\langle n_{\alpha}\sml r\smr\rangle$. The $\langle n_{\alpha}\sml r\smr\rangle$ are also directly calculable using the GMP state after a specific number of steps $r$ by evaluating
\begin{equation}
\label{3-29}
\langle n_{\alpha}\sml r\smr\rangle=\langle\Psi_{r}|\widehat{\mathbf{\Uppsi}}^{\dagger}{\sml {x}_{\alpha}\smr}\widehat{\mathbf{\Uppsi}}{\sml {x}_{\alpha}\smr}|\Psi_{r}\rangle=\sum_{\ell}\sum_{j} \bigg|\frac{C_{j\ell}^{r}}{\mathcal{K}_{r}}\bigg|^{2} n_{\alpha}.
\end{equation}
The Figs. \ref{fig:4}, \ref{fig:5} and \ref{fig:6}, present the expected vertices particle distribution $\langle n_{\alpha}\sml r\smr\rangle$ for a selected number of steps. We considered all three systems starting with an initial state of the form
 \begin{equation}
 \label{3-34}
|\Psi_{\mbox{\tiny{0}}}\rangle=\frac{C_{11}^{\mbox{\tiny{0}}}}{\mathcal{K}_{\mbox{\tiny{0}}}}|v_{\mbox{\tiny{1}}}{\bf n}_{\mbox{\tiny{1}}}\rangle+\frac{C_{22}^{\mbox{\tiny{0}}}}{\mathcal{K}_{\mbox{\tiny{0}}}}|v_{\mbox{\tiny{2}}}{\bf n}_{\mbox{\tiny{2}}}\rangle,
\end{equation}
 that is a linear combination of two configurations with the same amplitudes $\frac{C_{11}^{\mbox{\tiny{0}}}}{\mathcal{K}_{\mbox{\tiny{0}}}}={-i\over\sqrt{2}}$ and  $\frac{C_{22}^{\mbox{\tiny{0}}}}{\mathcal{K}_{\mbox{\tiny{0}}}}={1\over\sqrt{2}}$. We also considered $M=10$ vertices and $N=12$ particles initially distributed on a single vertex in two initial configurations: $|{\bf n}_{\mbox{\tiny{1}}}\rangle=|0,0,N,0,0,0,0,0,0,0\rangle$ and $|{\bf n}_{\mbox{\tiny{2}}}\rangle=|0,0,0,0,N,0,0,0,0,0\rangle$.
In addition to the vertices particle density, we also evaluate the phase space evolution, the second order spatial correlations and the vertices counting statistics. 
This scheme of many-particle quantum walks offers the possibility of accessing all types of correlations. The phase space is described by evaluating the position operator $\langle{x}_{\upeta}\sml r\smr\rangle=\langle\Psi_{r}|\hat{\bf x}_{\upeta}|\Psi_{r}\rangle$ 
\begin{equation}
\label{A-14.1}
\langle{x}_{\upeta}\sml r\smr\rangle=\sum_{\ell}\sum_{j}\bigg|\frac{C_{j\ell}^{r}(n_{\alpha})!}{\mathcal{K}_{r}n_{\alpha}^{n_{\alpha}}}\bigg|^{2}\sum_{\{{n}_{\upeta}\}}\prod_{\upeta}\cos[{\varphi}_{\upeta}{x}_{\alpha}],
\end{equation}
the momentum operator $\langle{p}_{\upeta}\sml r\smr\rangle=\langle\Psi_{r}|\hat{\bf p}_{\upeta}|\Psi_{r}\rangle$ 
\begin{equation}
\label{A-14.2}
\langle{p}_{\upeta}\sml r \smr\rangle=\sum_{\ell}\sum_{j}\bigg|\frac{C_{j\ell}^{r}(n_{\alpha})!}{\mathcal{K}_{r}n_{\alpha}^{n_{\alpha}}}\bigg|^{2}\sum_{\{{n}_{\upeta}\}}\prod_{\upeta}\sin[{\varphi}_{\upeta}{x}_{\alpha}],
\end{equation}
and the energy operator $\langle E_{\upeta}\sml r\smr\rangle=\langle\Psi_{r}|\hat{\bf E}_{\upeta}|\Psi_{r}\rangle$ 
\begin{eqnarray}
 \label{A-14.3}
 \langle{E}_{\upeta}\sml r\smr\rangle&=&\sum_{\ell}\sum_{j}\bigg|\frac{C_{j\ell}^{r}(n_{\alpha})!}{\mathcal{K}_{r}n_{\alpha}^{n_{\alpha}}}\bigg|^{2}\nonumber\\
 &&\quad\times\sum_{\{{n}_{\upeta}\}}\prod_{\upeta}\bigg({n}_{\upeta}+{1\over2}-\cos[2{\varphi}_{\upeta}{x}_{\alpha}]\bigg),
 \end{eqnarray}
calculated in each mode and at every step. These three quantities  in Eqs. (\ref{A-14.1}, \ref{A-14.2} and \ref{A-14.3}) represent the phase space coordinates in each mode. We recall that these operators are respectively defined by: $\hat{\bf x}_{\upeta}={1\over\sqrt{2}}(\hat{\bf b}_{\upeta}+\hat{\bf b}_{\upeta}^{\dagger})$, $\hat{\bf p}_{\upeta}={i\over\sqrt{2}}(\hat{\bf b}_{\upeta}-\hat{\bf b}_{\upeta}^{\dagger})$, and $\hat{\bf E}_{\upeta}={1\over2}\hat{\bf p}_{\upeta}^{\dagger}\hat{\bf p}_{\upeta}$. The constants such as the mass, the frequency and $\hbar$ are set to 1. The Figs. \ref{fig:7}, \ref{fig:8} and \ref{fig:9} present the results of 400 steps simulations where the color bar is the energy $\langle {E}_{\upeta}\sml r\smr\rangle$ in Eq \eqref{A-14.3}. 

Looking at the position second order correlations at step $r$, we use the spacial second order correlation function given by 
\vspace*{-0.25cm}
\begin{equation}
 \label{4-4.13}
 g_{(\alpha_{\mbox{\tiny{1}}},\alpha_{\mbox{\tiny{2}}},r)}^{\mbox{\tiny{(2)}}}=\frac{\langle\widehat{\mathbf{\Uppsi}}^{\dagger}{\sml {x}_{\alpha_{\mbox{\tiny{1}}}},r\smr}\widehat{\mathbf{\Uppsi}}^{\dagger} {\sml {x}_{\alpha_{\mbox{\tiny{2}}}},r\smr}\widehat{\mathbf{\Uppsi}}{\sml {x}_{\alpha_{\mbox{\tiny{2}}}},r\smr}\widehat{\mathbf{\Uppsi}}{\sml {x}_{\alpha_{\mbox{\tiny{1}}}},r\smr}\rangle}{\langle\widehat{\mathbf{\Uppsi}}^{\dagger} {\sml {x}_{\alpha_{\mbox{\tiny{2}}}},r\smr}\widehat{\mathbf{\Uppsi}}{\sml {x}_{\alpha_{\mbox{\tiny{2}}}},r\smr}\rangle\langle\widehat{\mathbf{\Uppsi}}^{\dagger} {\sml {x}_{\alpha_{\mbox{\tiny{1}}}},r\smr}\widehat{\mathbf{\Uppsi}}{\sml {x}_{\alpha_{\mbox{\tiny{1}}}},r\smr}\rangle}.
\end{equation}
These quantities are used to explore the influence of other vertices on a fixed vertex occupation number dynamics. 
The walks on the configurations Hilbert space are controlled by vertex occupation numbers. In Figs. \ref{fig:11} -- \ref{fig:43}, we present the results obtained for the $g_{(\alpha_{\mbox{\tiny{1}}},\alpha_{\mbox{\tiny{2}}},r)}^{\mbox{\tiny{(2)}}}$ for 400 steps of 12 quantum walkers over the three graphs in Fig. \ref{fig:123} where we present 
the $\mbox{30}^{\mbox{\tiny{th}}}$ step (Figs. \ref{fig:11} -- \ref{fig:13}), the $\mbox{50}^{\mbox{\tiny{th}}}$ step (Figs. \ref{fig:21} -- \ref{fig:23}), the $\mbox{100}^{\mbox{\tiny{th}}}$ step (Figs. \ref{fig:31} -- \ref{fig:33}) and the $\mbox{400}^{\mbox{\tiny{th}}}$ step (Figs. \ref{fig:41} -- \ref{fig:43}). 

\begin{figure*}[t]
  \centering
  \begin{subfigure}[b]{0.315\textwidth}
    \centering
    \includegraphics[height = 0.67\textwidth,width=1.1\textwidth]{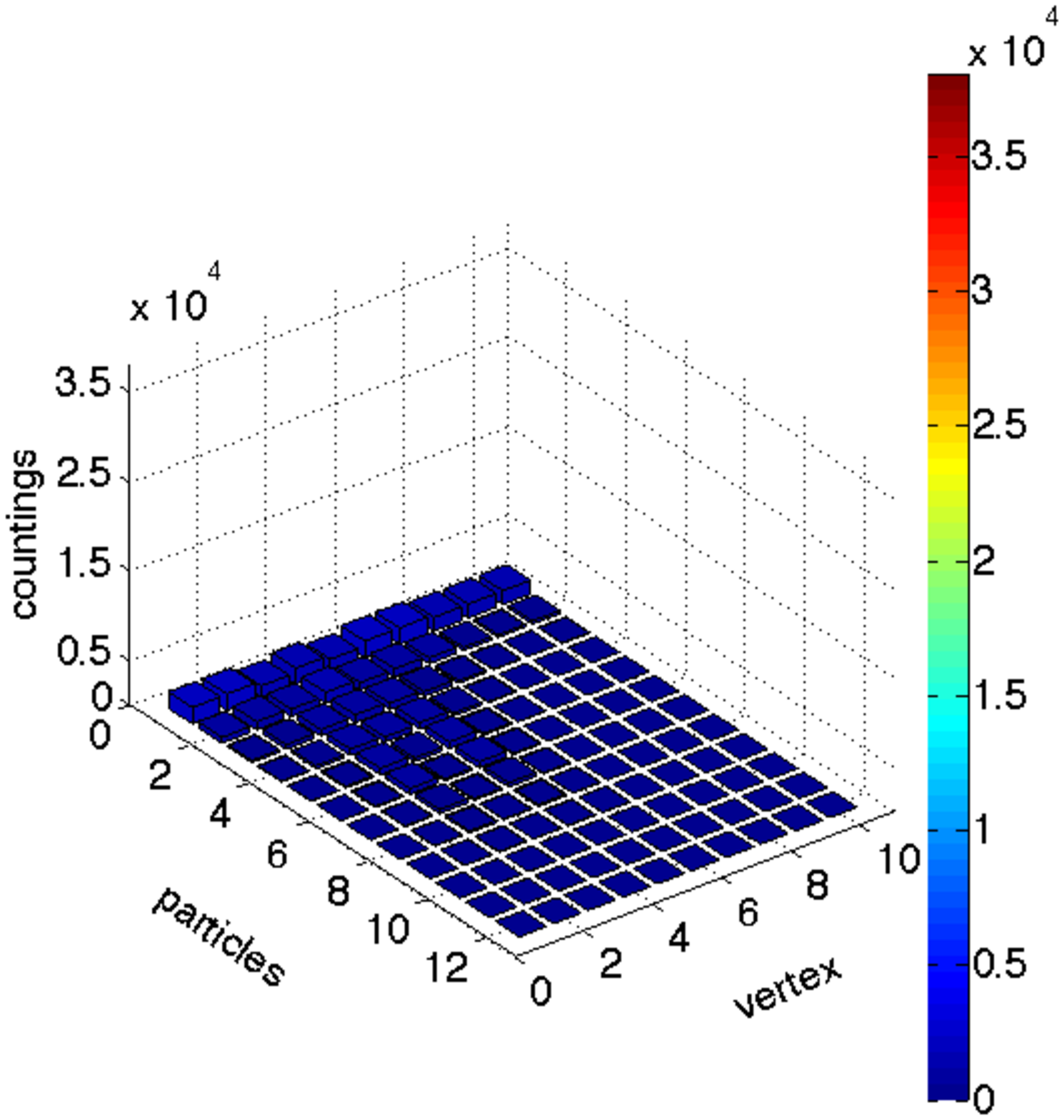}
    \caption{30 steps on cyclic graph\label{fig:51}}
  \end{subfigure}
  \hspace*{0.1cm}
  \begin{subfigure}[b]{0.315\textwidth}
    \centering
    \includegraphics[height = 0.67\textwidth,width=1.1\textwidth]{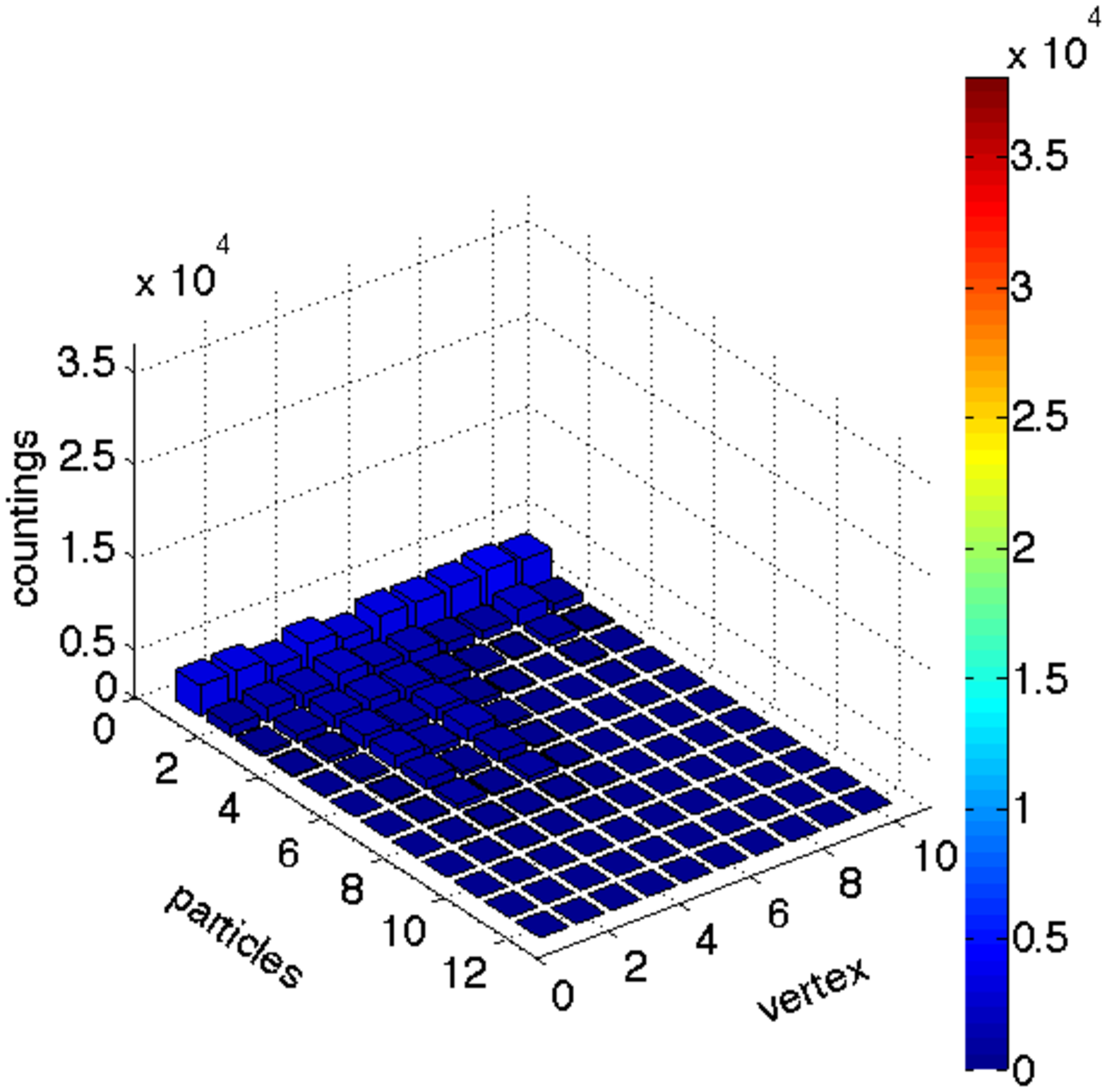}
    \caption{30 steps on double hexagon graph\label{fig:52}}
  \end{subfigure}
   \hspace*{0.1cm}
  \begin{subfigure}[b]{0.315\textwidth}
    \centering
    \includegraphics[height = 0.67\textwidth,width=1.1\textwidth]{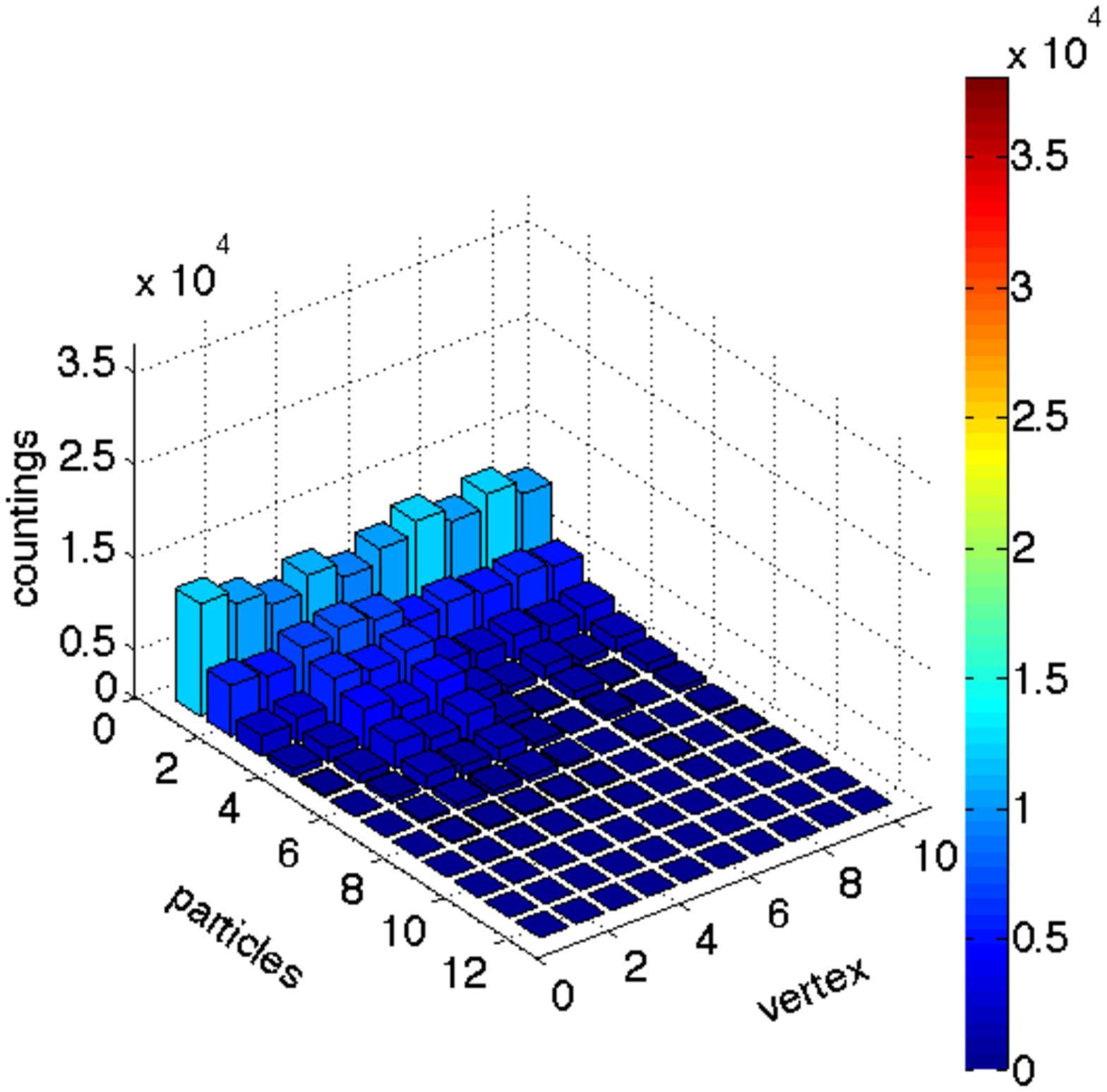}
    \caption{30 steps on Petersen graph\label{fig:53}}
  \end{subfigure}
  \begin{subfigure}[b]{0.315\textwidth}
    \centering
    \includegraphics[height = 0.67\textwidth,width=1.1\textwidth]{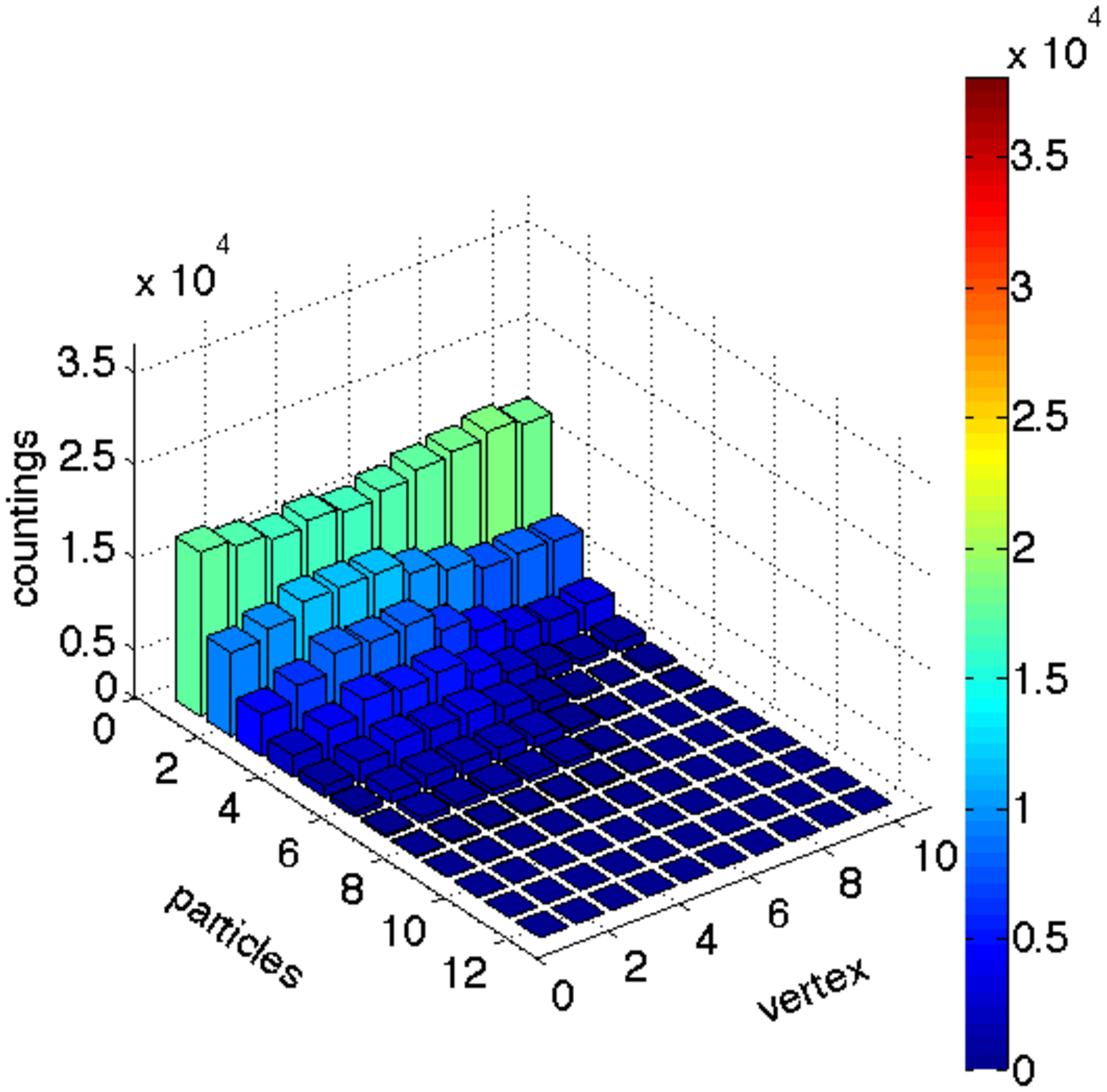}
    \caption{50 steps on cyclic graph\label{fig:61}}
  \end{subfigure}
  \hspace*{0.1cm}
  \begin{subfigure}[b]{0.315\textwidth}
    \centering
    \includegraphics[height = 0.67\textwidth,width=1.1\textwidth]{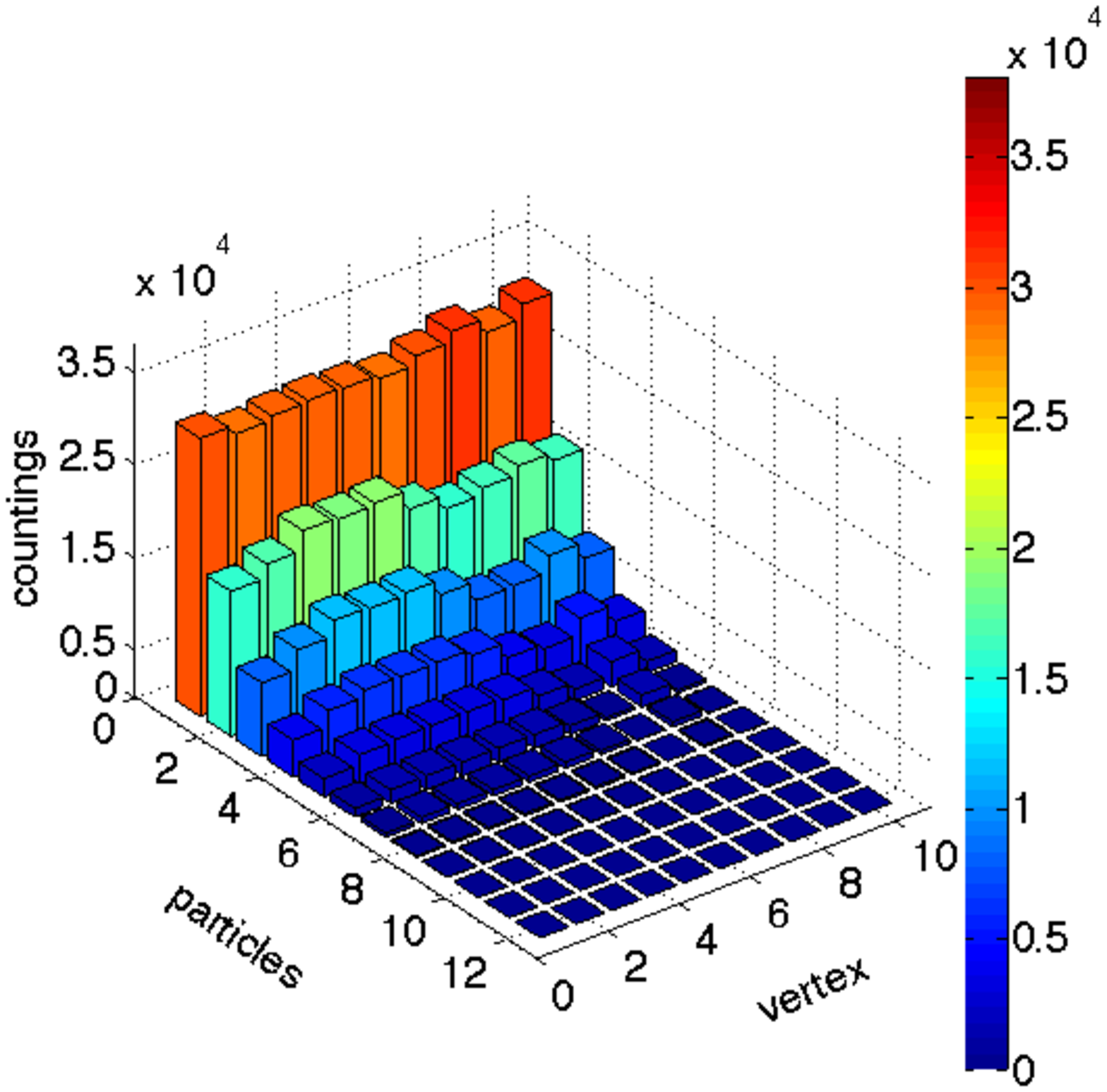}
    \caption{50 steps on double hexagon graph\label{fig:62}}
  \end{subfigure}
   \hspace*{0.1cm}
  \begin{subfigure}[b]{0.315\textwidth}
    \centering
    \includegraphics[height = 0.67\textwidth,width=1.1\textwidth]{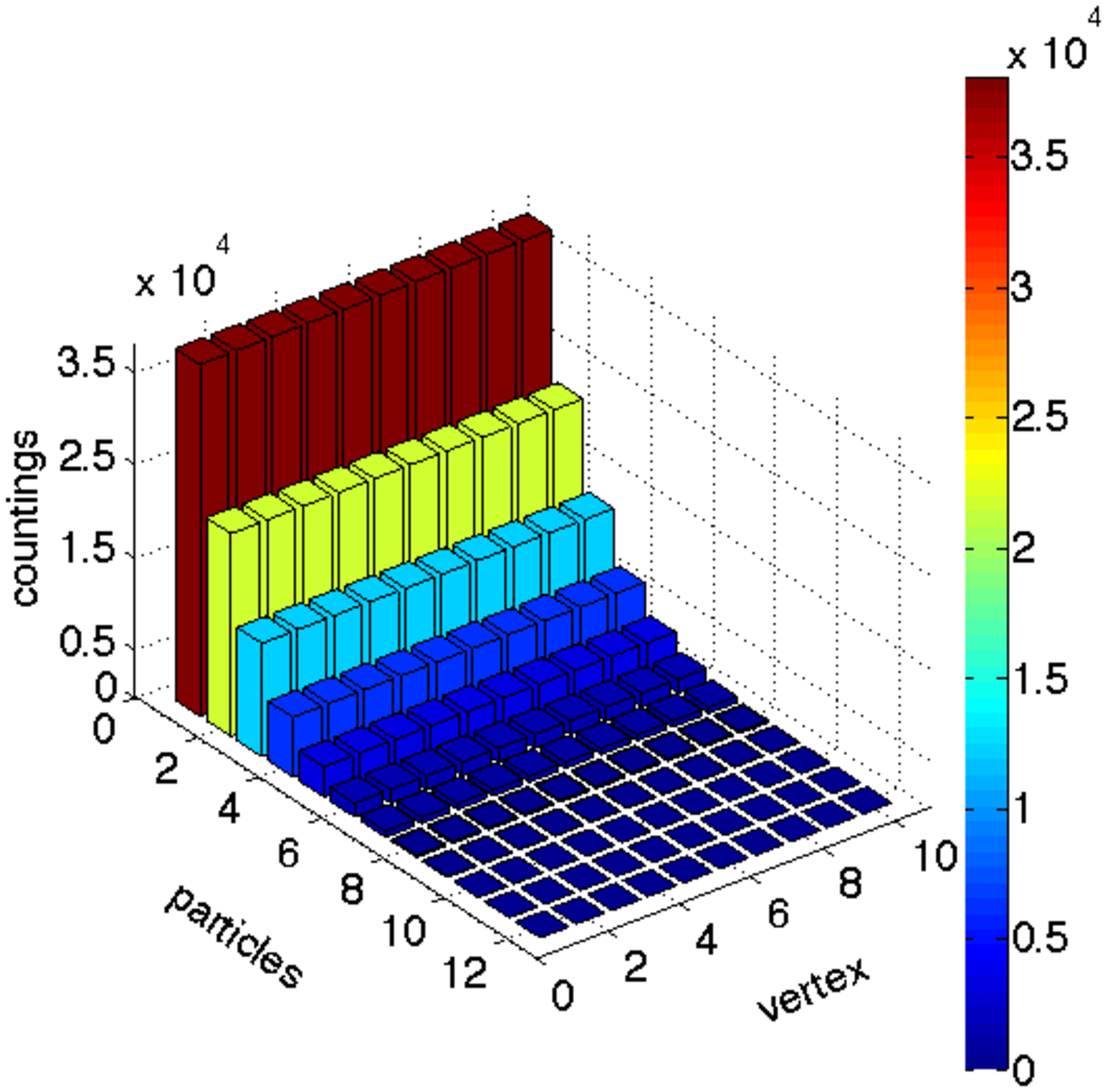}
    \caption{50 steps on Petersen graph\label{fig:63}}
  \end{subfigure}
  \begin{subfigure}[b]{0.315\textwidth}
    \centering
    \includegraphics[height = 0.67\textwidth,width=1.1\textwidth]{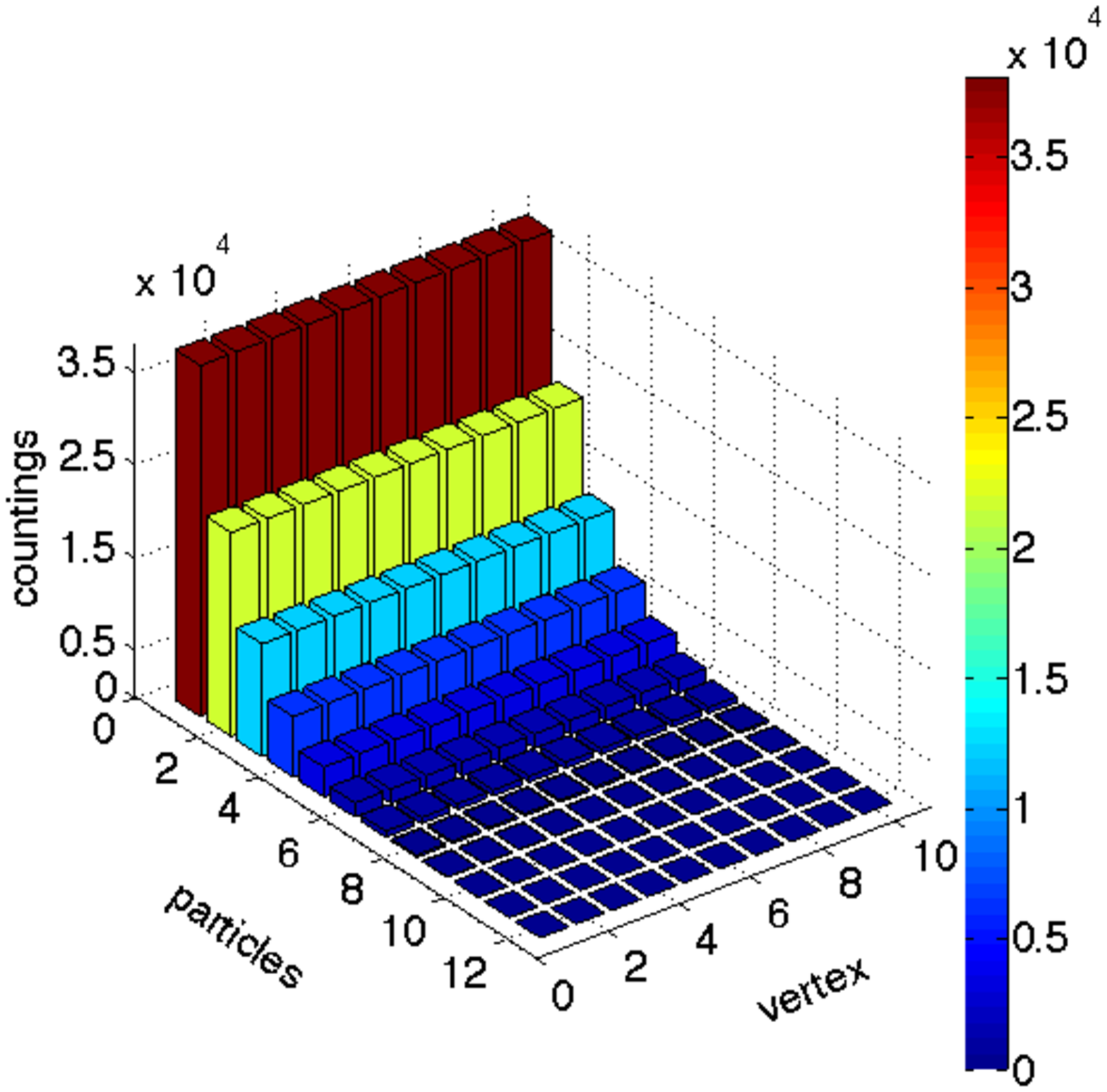}
    \caption{100 steps on cyclic graph\label{fig:71}}
  \end{subfigure}
  \hspace*{0.1cm}
  \begin{subfigure}[b]{0.315\textwidth}
    \centering
    \includegraphics[height = 0.67\textwidth,width=1.1\textwidth]{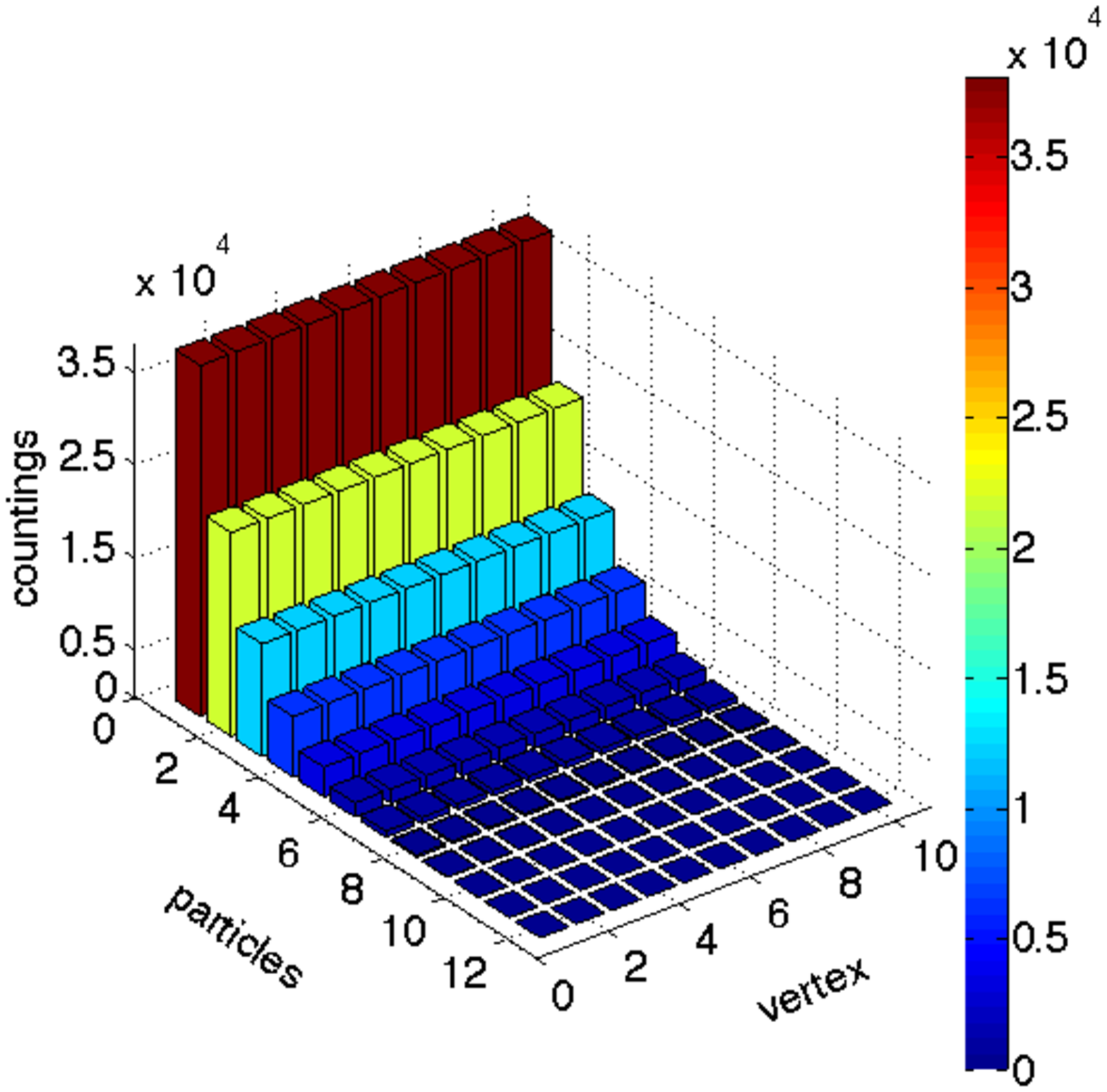}
    \caption{100 steps on double hexagon graph\label{fig:72}}
  \end{subfigure}
   \hspace*{0.1cm}
  \begin{subfigure}[b]{0.315\textwidth}
    \centering
    \includegraphics[height = 0.67\textwidth,width=1.1\textwidth]{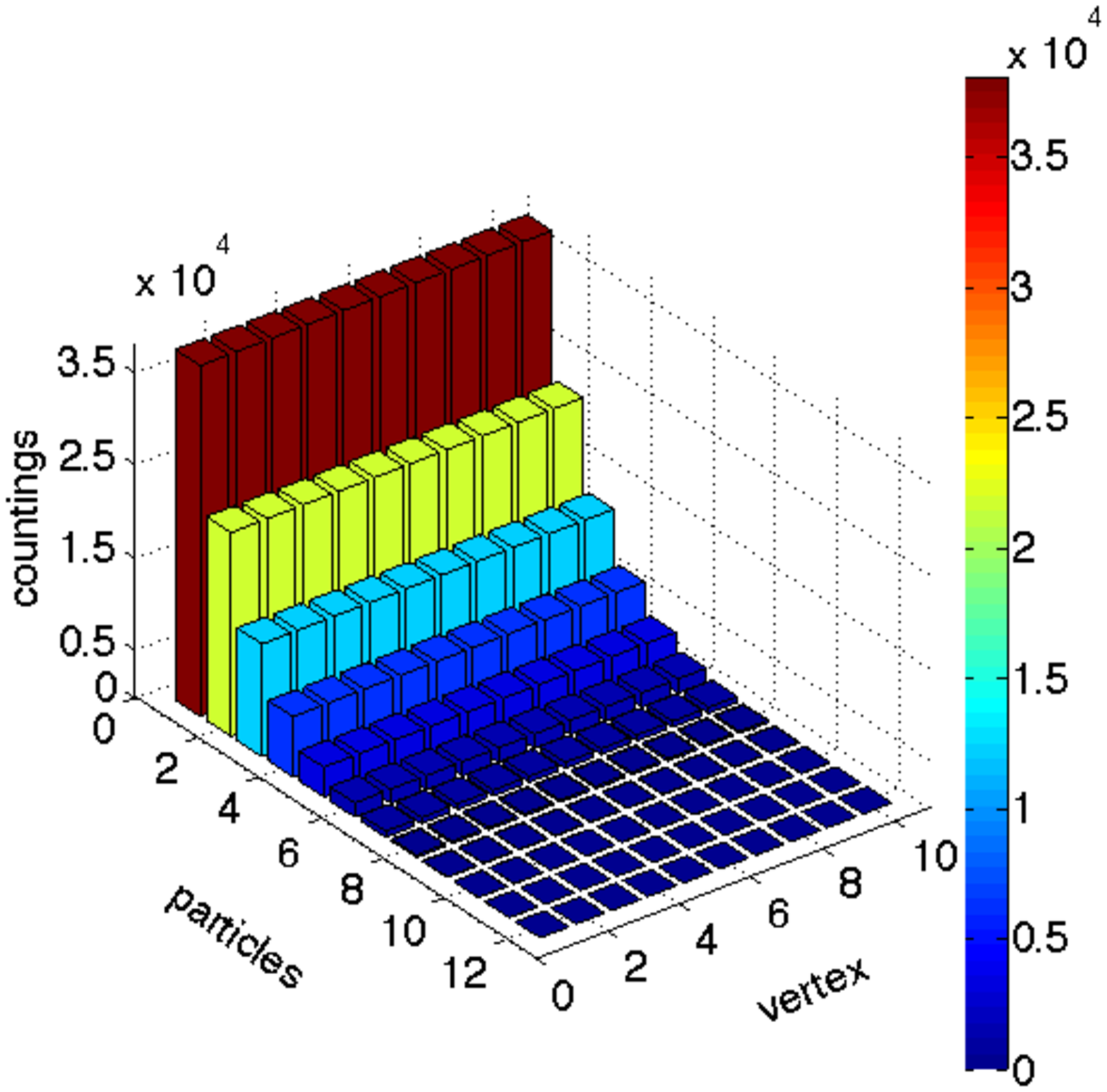}
    \caption{100 steps on Petersen graph\label{fig:73}}
  \end{subfigure}
  \begin{subfigure}[b]{0.315\textwidth}
    \centering
    \includegraphics[height = 0.67\textwidth,width=1.1\textwidth]{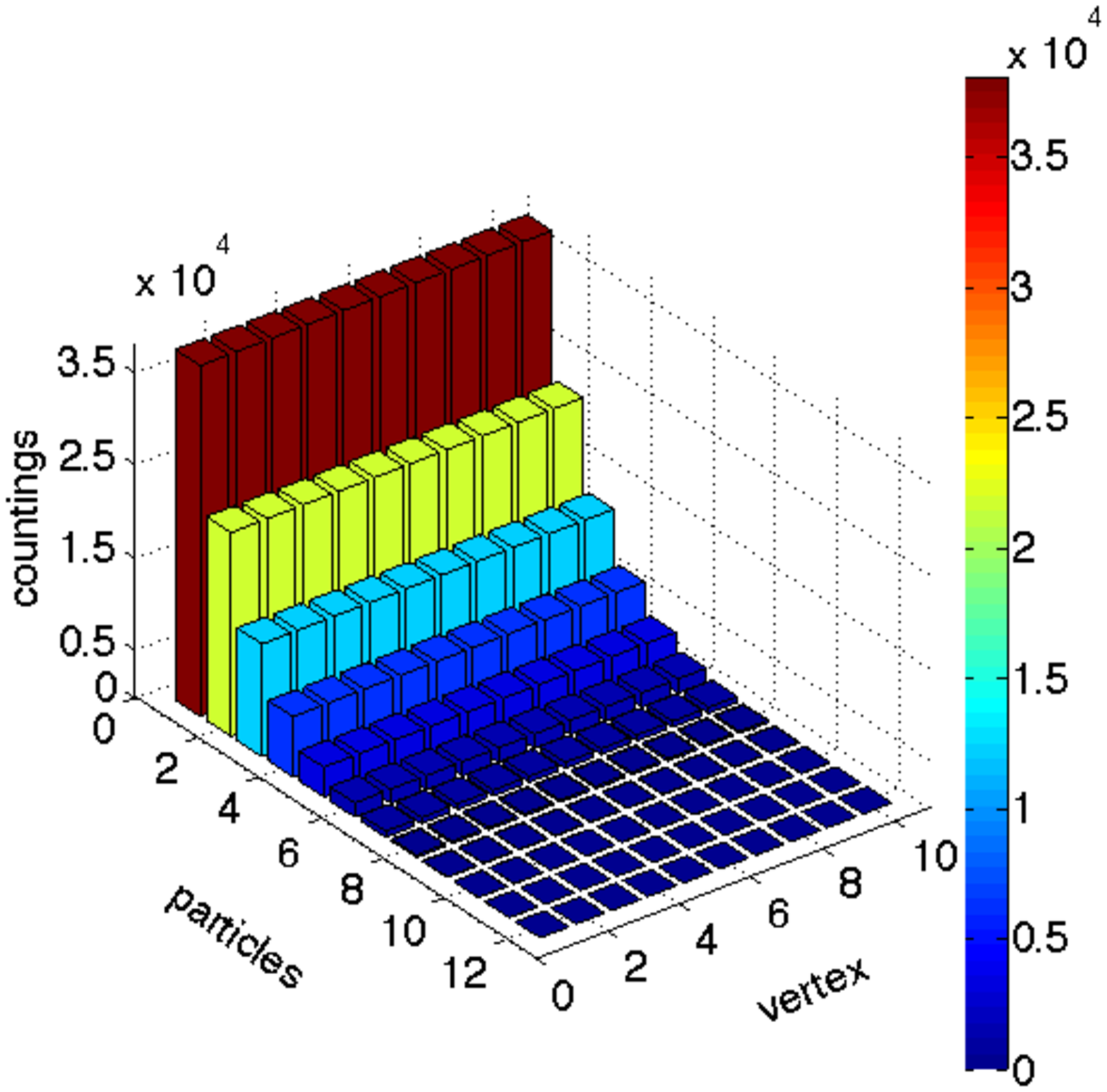}
    \caption{400 steps on cyclic graph\label{fig:81}}
  \end{subfigure}
  \hspace*{0.1cm}
  \begin{subfigure}[b]{0.315\textwidth}
    \centering
    \includegraphics[height = 0.67\textwidth,width=1.1\textwidth]{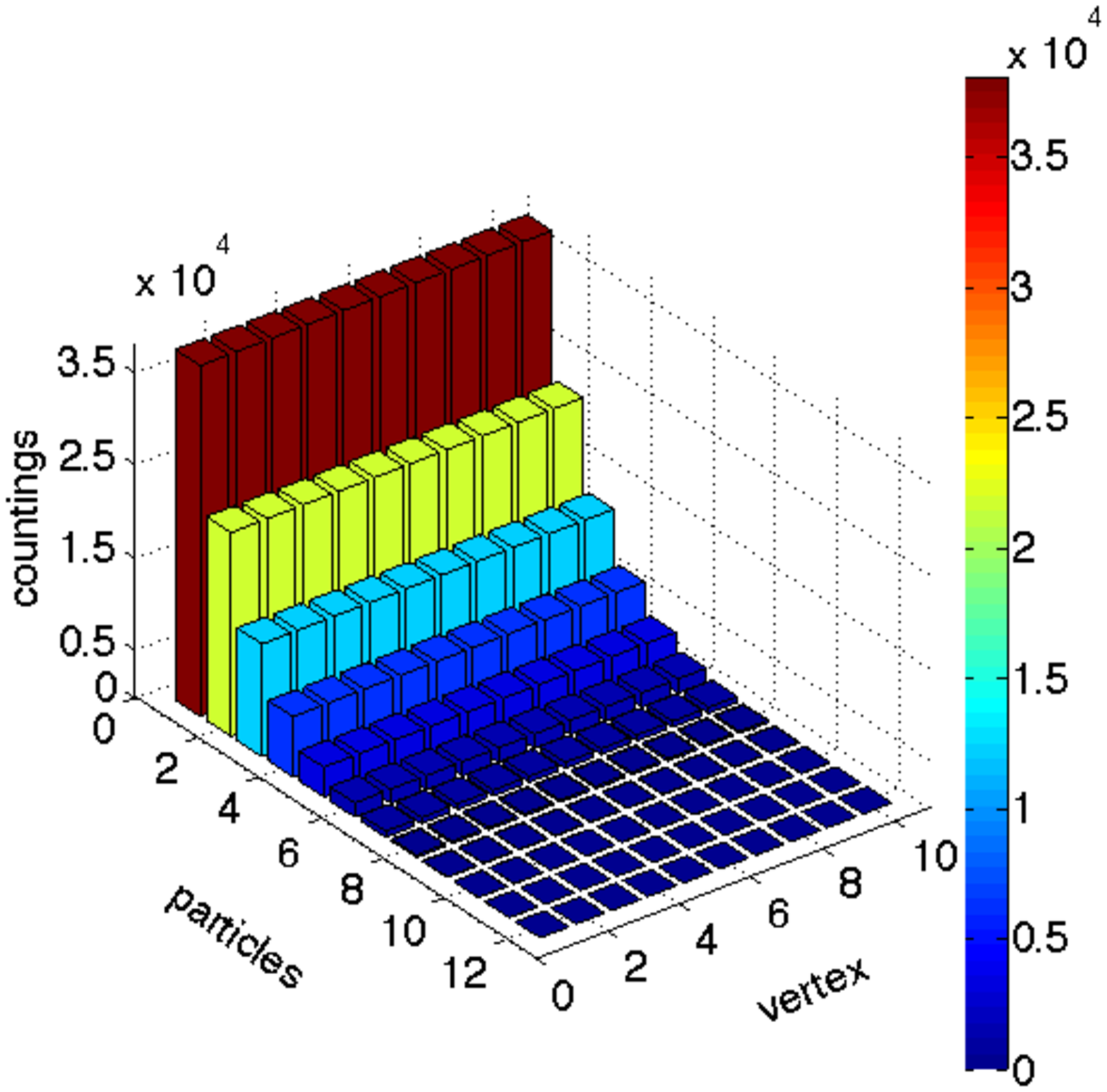}
    \caption{400 steps on double hexagon graph\label{fig:82}}
  \end{subfigure}
   \hspace*{0.1cm}
  \begin{subfigure}[b]{0.315\textwidth}
    \centering
    \includegraphics[height = 0.67\textwidth,width=1.1\textwidth]{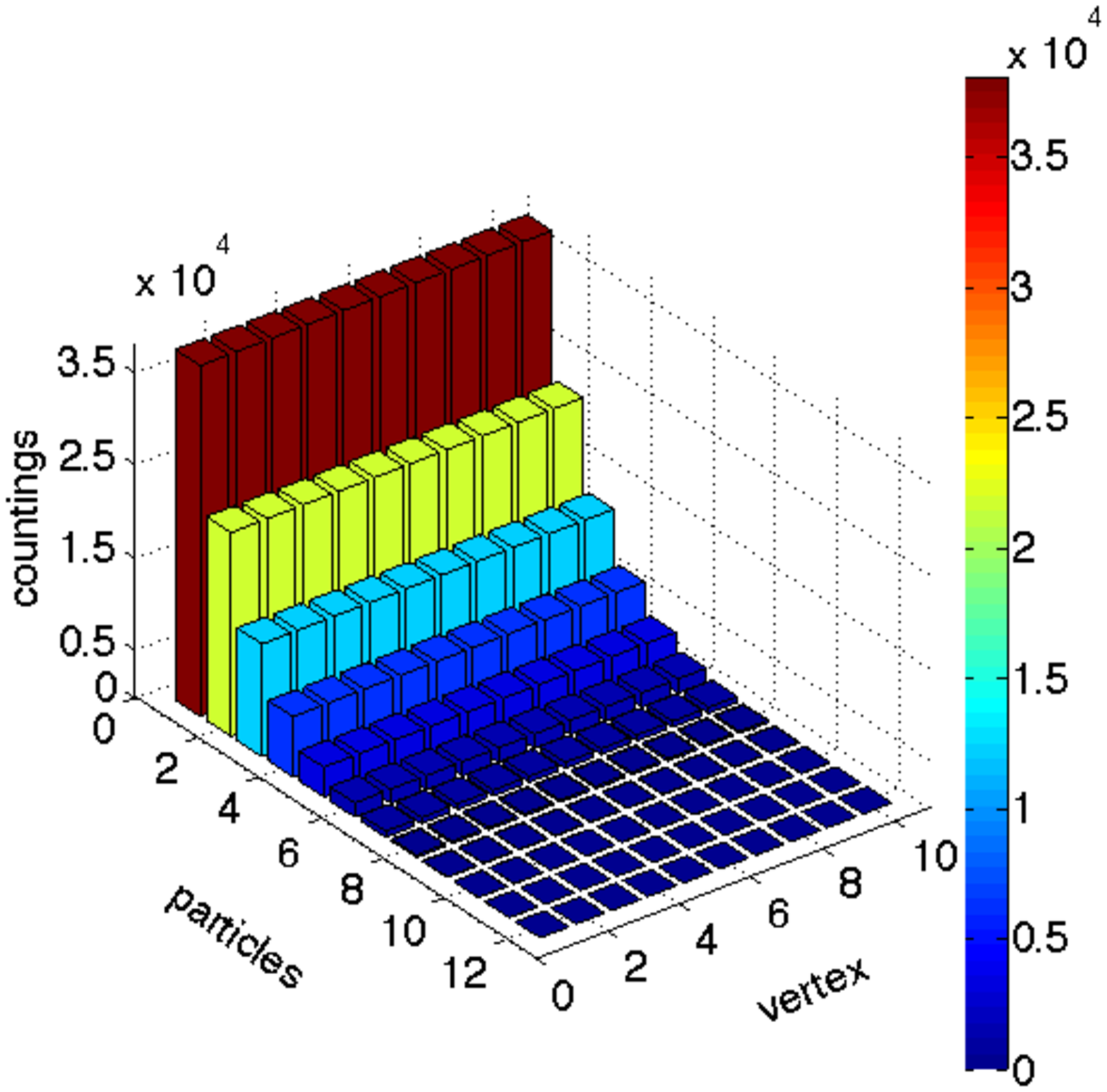}
    \caption{400 steps on Petersen graph\label{fig:83}}
  \end{subfigure}
   \caption{ Selected steps simulations representing the vertices counting statistics of 12 quantum walkers on 10 vertices graphs. Before the change of regime we have:  (a)--(d) for the cyclic graph, (b)--(e)  for the double hexagon graph and (c)  for the Petersen graph.  After the change of regime we have: (g)--(j) for the  cyclic graph, (h)--(k)  for the double hexagon graph and (f)--(i)--(l) for the Petersen graph.}
\vspace*{-0.2cm}
\end{figure*}

There are various methods of exploring the evolution of the configurations Hilbert space. In this case we are focusing on microscopic evolutions by looking at a specific vertex $\alpha$. To such a vertex we join a counting $n_{\alpha}$. This is equivalent to fine tuning the detector to record a count only when $n_{\alpha}$ walkers report on vertex $\alpha$. This counting combines the combinatoric problem  of distributing of $N$ indistinguishable bosons into $M$ vertices to the probability of a given configuration as in Eq. \eqref{3-27}. The probability of finding $n_{\alpha}$ bosons on vertex $\alpha$ considered as the combinatorial problem of distributing of $N$ indistinguishable bosons into $M$ vertices is given by: ${D\tyl \mbox{\tiny{$N$}}-n_{\alpha},\mbox{\tiny{$M$}}\tyr\over{MD\tyl \mbox{\tiny{$N,M$}}\tyr}}$ where $D\tyl\cdot,\cdot\tyr$ is defined in Eq.\eqref{1}
We define the probability $P_{n_{\alpha}}^{r}$ of finding $n_{\alpha_{\mbox{\tiny{1}}}}=n_{\alpha}$ particles on vertex $\alpha$ at step $r$ as: 
\begin{equation}
\label{2-4.50.1}
P_{n_{\alpha}}^{r}=\sum_{j}\frac{(C_{j\ell}^{r})^*C_{j\ell}^{r}}{[\mathcal{K}_{r}]^2M}\frac{D\sml N-n_{\alpha},M-1\smr}{D\sml N,M\smr}.
\end{equation}
Considering the evolution of vertex by vertex counting statistics we observe a change of regime in the time step evolution counting statistics. This change appears in all the three graphs but at different time steps. For example for all these three systems, when the quantum walks start from the initial conditions in Eq.\eqref{3-34} the change of regime occurs after $r=94$ steps for the cyclic graph in Fig. \ref{fig:123}a, after $r=70$ steps for the double hexagon graph in Fig. \ref{fig:123}b and after $r=48$ steps for the Petersen graph in Fig. \ref{fig:123}c. Before the change of regime the number of the counts grows quickly specially for the small number of particles on a vertex see Figs. \ref{fig:51}--\ref{fig:61}, Figs. \ref{fig:52}--\ref{fig:62} and Figs. \ref{fig:53}--\ref{fig:63}.  After the change of regime the counting statistics remains consistently the same for all the three graphs and for different initial conditions. During our simulations we also monitored what fraction of the configuration Hilbert space was effectively contributing in the GMP state with nonzero amplitudes. Let us simply name it the effective Hilbert space. Moreover, the observation of regime change in the counting statistics time evolution corresponds  with the change of behavior of the dimension of the effective Hilbert space. Before the change of the regime this dimension grows quickly e.g for the cyclic graph from $7900$ at step $r=30$ to $68632$ at step $r=50$ (Corresponding numbers for the double hexagon graph are: $14507$, $115052$). At the change of regime step, the dimension of the effective Hilbert space reaches one of the  two critical values $146860$ or  $147070$ and at the subsequent steps oscillates between these two values. We say that this behavior is universal because it is present in different types of graphs and for different initial conditions as long as the number of vertices over the considered graphs is the same and the number quantum walkers remains the same.  In this work we have present only the results
for one initial condition in  Eq. \eqref{3-34} shown in Figs. \ref{fig:51}-\ref{fig:83}. We observed the same phenomenon for other initial conditions. It means that the vertices counting statistics shows universal behavior after the change of regime step for each system. It must also be noted that the change of regime step depends on the initial conditions for the designated graph in Fig.\ref{fig:123}. 

Phenomenologically it can be observed that during the change of regime, the systems enter into a two degrees of freedom phase. In other words, during successive steps, the GMP states in Eq. \eqref{3-12.2} related to the three graphs evolve between two subspaces of the configuration Hilbert space of dimensions 146860 and 147070, respectively. Looking at generic many-particle quantum systems, it is conjectured that thermalization is possible at the level of individual eigenstates \cite{Deutsch, Srednicki}. Rigol et al in works \cite{Rigol01,Rigol03,Rigol02} on thermalization in generic isolated quantum systems suggest that despite the fact that an isolated quantum system doesn't thermalize like a classical system, it is possible to identify an observable for which the mean values behave such as it is predicted using an appropriate statistical mechanical ensemble (represented by a certain density matrix). Authors of \cite{Rigol02} use the evolution of particles-mode distribution $n_{\upeta}$ to monitor numerically the thermal relaxation. In this work, we have chosen the vertices counting statistics to monitor the time evolution of the system. Under the hypothesis of eigenstate thermalization, the system evolves from an initial state with non-thermalized eigenstates to a final state with thermalized eigenstates. Therefore it is expected that individual many-particle eigenstates thermalize without affecting the dimension of the effective configuration Hilbert space. Consequently, under the hypothesis of eigenstates thermalization the GMP state during the shared coins many-particle quantum walks on closed graphs evolves toward a subspace of the configuration Hilbert space called effective Hilbert space of desired dimension. However, we are observing a cyclic changes of the dimension of the effective configuration Hilbert space for systems on graphs in Figs. \ref{fig:123} where some eigenstates coalesce during the evolution from the effective Hilbert space of dimension $147070$ to the effective Hilbert space of dimension $146860$ and split during the reverse evolution.  

Moreover, we notice that the dimension of the effective Hilbert space oscillates between values that are roughly a third of dimension of the configurations Hilbert space.  Then under the hypothesis of ergodicity, there must exist additional conserved physical observables including the Hamiltonian that are functionally independent each other. In such a case the evolution of the system will be restricted to a subspace of the configurations Hilbert space and the uniform distribution for the many-particle quantum walks becomes inaccessible. Such cases have been experimentally observed like in Fermi-Pasta-Ulam numerical experiment \cite{FerUlam01}. Nevertheless, looking at the evolution of the second order spacial correlations in Figs.\ref{fig:41}--\ref{fig:43}, one can observe that the quantum walks have a finite confinement time and they are slowly evolving toward a uniform distribution in effective Hilbert space, see Fig \ref{fig:6}.  In conclusion, we conjecture that these systems become integrable \cite{Toshiya01,JHarnard01, Izergin:01} after the regime change steps. To understand the limits of this conjecture, it is useful to explore other observables involved in the dynamics. Here we have in mind higher order correlations that might be of relevance as well as other observables directly connected to the microscopic evolution of the configurations Hilbert space. \\
This research was supported by grant No. DEC-2011/02/A/ST1/00208 of National Science Centre of Poland.
\vspace*{-0.4cm}
%\bibliographystyle{apsrev4-1}
%\bibliographystyle{plain}
%\bibliography{pierrot,pierrotmanyqwalksNotes}

\end{document}